\documentclass[pra,twocolumn,floatfix,superscriptaddress]{revtex4}

\usepackage{graphicx}
\usepackage{color}
\usepackage{subfigure}
\usepackage{tikz}
\usetikzlibrary{automata,calc,backgrounds,arrows,positioning, shapes.misc,quotes,arrows.meta}
\usepackage[normalem]{ulem}
\usepackage{braket}
\usepackage{amsmath}
\usepackage{verbatim}
\usepackage{amssymb}
\usepackage{mathtools}
\usepackage{bm}

\usepackage[pdfstartview=FitH]{hyperref}
\hypersetup{
    colorlinks=true,       % false: boxed links; true: colored links
    linkcolor=cyan,          % color of internal links
    citecolor=magenta,        % color of links to bibliography
    filecolor=magenta,      % color of file links
    urlcolor=cyan,           % color of external links
    runcolor=cyan
}

\newcommand{\beq}{\begin{equation}}
\newcommand{\eeq}{\end{equation}}
\newcommand{\bea}{\begin{eqnarray}}
\newcommand{\eea}{\end{eqnarray}}
\newcommand{\nn}{\nonumber \\}
\newcommand{\E}{\mathbb{E}}
\newcommand{\Ham}{\mathcal{H}}
\newcommand{\z}{\mathcal{\bzeta}}
\def\pd{\partial} 

\def\x{\mathbf{x}}
\def\z{\mathbf{z}}
\def\bzeta{\bm{\zeta}}
\def\btheta{{\bm{\theta}}}

\def\brho{{\bm{\rho}}}
\def\bphi{{\bm{\phi}}}

\def\tr{\text{Tr}}
\def\Ex{\mathbb{E}_{\x \sim p_\text{data}}}

\begin{document}

\title{A Path Towards Quantum Advantage in \\ Training Deep Generative Models with Quantum Annealers}

\author{Walter Vinci}
\affiliation{Quantum Artificial Intelligence Lab. (QuAIL), Exploration Technology Directorate, NASA Ames Research Center, Moffett Field, CA 94035, USA}
\affiliation
{Stinger Ghaffarian Technologies Inc., Greenbelt, MD 20770, USA}

\author{Lorenzo Buffoni}
\affiliation
{D-Wave Systems Inc., 3033 Beta Avenue, Burnaby BC
Canada V5G 4M9}
\affiliation
{Department of Physics and Astronomy, University of Florence, via G. Sansone 1, I-50019 Sesto Fiorentino, Italy}
\affiliation{Department of Information Engineering, University of Florence, via S. Marta 3, I-50139 Florence, Italy}

\author{Hossein Sadeghi}
\affiliation
{D-Wave Systems Inc., 3033 Beta Avenue, Burnaby BC
Canada V5G 4M9}

\author{Amir Khoshaman}
\affiliation{D-Wave Systems Inc., 3033 Beta Avenue, Burnaby BC
Canada V5G 4M9}

\author{Evgeny Andriyash}
\affiliation
{D-Wave Systems Inc., 3033 Beta Avenue, Burnaby BC
Canada V5G 4M9}

\author{Mohammad H. Amin }
\affiliation{D-Wave Systems Inc., 3033 Beta Avenue, Burnaby BC
Canada V5G 4M9}
\affiliation{Department of Physics, Simon Fraser
University, Burnaby, BC Canada V5A 1S6}

\begin{abstract}
The development of quantum-classical hybrid (QCH) algorithms is critical to achieve state-of-the-art computational models. A QCH variational autoencoder (QVAE) was introduced in Ref.~\cite{khoshaman2018quantum} by some of the authors of this paper. QVAE consists of a classical auto-encoding structure realized by traditional deep neural networks to perform inference to, and generation from, a discrete latent space. The latent generative process is formalized as thermal sampling from either a quantum or classical Boltzmann machine (QBM or BM). This setup allows quantum-assisted training of deep generative models by physically simulating the generative process with quantum annealers. In this paper, we have successfully employed D-Wave quantum annealers as Boltzmann samplers to perform quantum-assisted, end-to-end training of QVAE. The hybrid structure of QVAE allows us to deploy current-generation quantum annealers in QCH generative models to achieve competitive performance on datasets such as MNIST. The results presented in this paper suggest that commercially available quantum annealers can be deployed, in conjunction with well-crafted classical deep neutral networks, to achieve competitive results in unsupervised and semisupervised tasks on large-scale datasets. We also provide evidence that our setup is able to exploit large latent-space (Q)BMs, which develop slowly mixing modes. This expressive latent space results in slow and inefficient classical sampling, and paves the way to achieve quantum advantage with quantum annealing in realistic sampling applications.

\end{abstract}

\maketitle
% %%%%%%%%%%%%%%%%%%%%%%%%%%%%%%%%%%%%%
% \section{Introduction}
% \label{sec:intro}
% %%%%%%%%%%%%%%%%%%%%%%%%%%%%%%%%%%%%%

Deep learning~\cite{rosenblatt1958perceptron,rumelhart1988learning,hinton2006fast,bengio2007greedy,lecun2015deep}, in which a labeled dataset is used to train a statistical model to solve tasks such as clustering and classification, is currently revolutionizing the field of supervised learning. Deep neural networks are now  commonly used in many scientific and industrial applications. Unlike supervised learning~\cite{goodfellow2016deep}, unsupervised learning is a much harder, and still largely unsolved, problem. And yet, it has the appealing potential to learn the hidden statistical correlations of large unlabeled datasets~\cite{vincent2008extracting,hinton1995wake,bengio2007greedy}, which constitute the vast majority of data available today. 

Training and deployment of large-scale machine learning models, especially for unsupervised learning, faces computational challenges~\cite{tieleman2008training} that are only partially met by the development of special purpose classical computing units such as GPUs. This has led to an interest in applying quantum computing to machine learning tasks~\cite{schuld2015introduction,wittek2014quantum,adcock2015advances,arunachalam2017survey,Biamonte:2017db} and to the development of several quantum algorithms~\cite{harrow2009quantum,wiebe2012quantum,childs2015quantum,lloyd2014quantum} with the potential to accelerate training. Most quantum machine learning algorithms need fault-tolerant quantum computation~\cite{nielsen2002quantum,lidar2013quantum,fowler2012surface}, which requires the large-scale integration of millions of qubits and is still not available today. It is however believed that quantum machine learning (QML) will provide the first breakthrough algorithms to be implemented on commercially available quantum annealers~\cite{farhi2001quantum,johnson2011quantum} and gate-model devices~\cite{neill2017blueprint,kandala2017hardware}. For example, small gate-model devices and quantum annealers have been used to perform quantum heuristic optimization~\cite{kadowaki1998quantum,santoro2002theory,brooke2001tunable,farhi2014quantum,peruzzo2014variational,kandala2017hardware} to solve clustering~\cite{otterbach2017unsupervised} and classification problems~\cite{neven2008training,denchev2012robust,pudenz2013quantum,mott2017solving}.

Current-generation quantum annealers physically simulate a transverse-field Ising model and operate in interaction with in a thermal environment. Perhaps the most natural use of quantum annealers is thus  as samplers from the Boltzmann distribution of an Ising system~\cite{amin2015searching,venuti2016adiabaticity,albash2017temperature,vinci2017scalable}. This observation inspired the use of quantum annealers for the training of Boltzmann machines (BMs) ~\cite{denil2011toward,raymond2016global,korenkevych2016benchmarking,perdomo2017opportunities,adachi2015application,benedetti2016estimation,benedetti2016quantum},  powerful and versatile probabilistic models that can be trained in either a supervised or unsupervised fashion. Training BMs  requires computationally-expensive techniques, such as persistent contrastive divergence (PCD)~\cite{tieleman2008training} and population annealing (PA)~\cite{hukushima2003population}, to accurately sample from the correct thermal distribution. Quantum annealing has the potential to overcome this computational bottleneck and to allow training of larger and more powerful BMs. Additionally, quantum annealers with advanced annealing control schedules allow sampling from quantum BMs (QBMs)~\cite{Amin:2016qd, harris2018phase, king2018observation}, which might be able to capture the more complex correlations realized by quantum states. Despite being able to represent complex probability distributions, early numerical studies showed that restricted BMs (RBMs) with the quasi two-dimensional connectivities typical of current-generation quantum annealers perform poorly~\cite{dumoulin2014challenges}, even on simple standard datasets such as MNIST~\cite{lecun1998mnist}.

A possible approach to circumvent the limitation above is the following quantum-classical hybrid (QCH) paradigm: 1) employ classical feed-forward neural networks (NN) to encode the  data into a compressed representation that is more easily processed by a quantum device and 2) jointly train the NN and the quantum device.  This approach was pioneered in Ref.~\cite{benedetti2018quantum}, where the authors introduced a quantum-assisted Helmholtz machine (QAHM), a generative model with latent variables. A QAHM has an \emph{inference} (encoder), a \emph{generation} (decoder) network and a generative process that is represented by a QBM or classical BM, and can thus be simulated by a quantum annealer. QAHMs do not have a well-defined loss function~\cite{hinton1995wake}, which means that training does not scale well, even to standard machine-learning datasets such as MNIST. This approach was taken a step further in Ref.~\cite{khoshaman2018quantum}  by using variational autoencoders (VAE), a class of generative models with latent variables that provide an efficient inference mechanism~\cite{kingma2013auto,burda2015importance}. VAEs can be trained by minimizing the evidence lower bound (ELBO), a variational lower  bound to the exact log-likelihood. The ELBO is a well-defined loss function with fully propagating gradients that can be efficiently optimized via backpropagation. This allows us to achieve competitive and scalable performance on  large-scale datasets. Upon discretization of the latent space~\cite{rolfe2016discrete,khoshaman2018gumbolt}, the generative process can be then realized by either classical or quantum BMs. In the latter case, training can be also performed efficiently by optimizing a quantum lower  bound to the exact log-likelihood~\cite{khoshaman2018quantum}. 

In this paper we use quantum annealers to sample from a classical BM. In other words, we assume that the quantum bias due to the presence of a transverse field is small enough that we can use a quantum annealer to approximate thermal expectations of a classical BM. The thermal expectation of the BM energy (the ``negative phase") is indeed required to compute the gradients of the BM parameters~\cite{ackley1985learning}. We demonstrate that the evaluation of these thermal expectations with a quantum annealer is accurate enough to allow the training of a VAE with latent-space BM. 

Our hybrid approach exploits a large amount of classical computational resources (backpropagation through deep NN performed on GPUs). This is advantageous since it allows us to achieve competitive results on relatively large-scale datasets such as MNIST. However, validating the training with quantum annealers is a major challenge: VAE can train well even with unstructured priors (\emph{e.g.}, a product of Gaussian distributions). One thus needs to carefully validate and demonstrate any performance improvement due to the use of structured samples coming from well-trained BMs. To this end, existence of a well-defined loss function, which allows for a quantitative comparison between different models, is critical. In order to single out the performance improvement due to computations that can be off-loaded to a quantum annealer (sampling from the BM), we always focus on comparing the performance of a model trained with a nontrivially connected BM (for example with full or Chimera connectivity \ref{sec:CHI-PEG}) to a fully classical baseline in which we employ a BM with no connectivity; that is, a set of independent Bernoulli variables. After a certain number of gradient updates, training of such models will have used exactly the same amount of classical resources that cannot be off-loaded to the quantum annealer, with the only difference being the computational effort in sampling from the latent-space BM.

We successfully train a convolutional VAE with 288 latent units and Chimera-structured RBM (a 6-by-6 Chimera graph; see Appendix~\ref{sec:CHI-PEG}) using only samples obtained by D-Wave 2000Q quantum annealers to estimate the negative phases required to evaluate the gradients for the RBM parameters. We then use several techniques to validate the trained models. In particular, we use an auxiliary RBM trained on the annealer samples to quantitatively estimate the test log-likelihood and compute several model parameters to show how they track the same quantities for models trained fully classically. With these approaches, we show that models trained on Chimera connectivity outperform their trivial (``Bernoulli") baseline and can achieve competitive performance on the MNIST dataset.

The next question is whether our approach offers a path towards obtaining quantum advantage with quantum annealers in machine learning applications. To achieve such a goal, we need to exploit large latent-space RBMs that develop complex multimodal probability distributions (\emph{i.e.}, a complex energy landscape) from which sampling is classically inefficient. We give evidence that this is indeed possible. For example, we demonstrate that the BMs placed in the latent space develop nontrivial modes that are likely to cause classical Monte Carlo algorithms to have long mixing times. Moreover, we show that training on more complex datasets likely takes advantage of larger BMs to improve performance. In addition, we discuss the role of connectivity, emphasizing its importance even in this hybrid approach, and the necessity to develop device-specific classical NN to better exploit physical connectivities such as the Chimera graph. 

The structure of the paper is as follows. In Sec.~\ref{sec:QVAE} we review VAE and the implementations of discrete latent variables, a necessary step to implement BMs and QBMs in their latent space. Section \ref{sec:QA} motivates the use of quantum annealers as samplers to train quantum and classical BMs. In Sec.~\ref{sec:HW} we report our experiments in training VAEs with D-Wave 2000Q systems. In Sec.~\ref{sec:path} we discuss a possible path toward quantum advantage in our setup. Finally, we present our conclusions in Sec.~\ref{sec:conc}.

%%%%%%%%%%%%%%%%%%%%%%%%%%%%%%%%%%%%%
\section{Variational Autoencoders}
\label{sec:QVAE}
%%%%%%%%%%%%%%%%%%%%%%%%%%%%%%%%%%%%%
In this section, we will briefly introduce VAEs and describe their extension to discrete latent variables; a necessary step to hybridize with quantum priors and to perform quantum-assisted training.

% %%%%%%%%%%%%%%%%%%%%%%%%%%%%%%%%%%%%%
% \subsection{Variational autoencoders}
% %%%%%%%%%%%%%%%%%%%%%%%%%%%%%%%%%%%%%

\begin{figure}[t]
\begin{center}
\begin{tikzpicture}[->,>=stealth',shorten >=1pt,auto,node distance=1.0cm, thick, scale=1]
       \tikzstyle{every state}=[fill=white,draw=black,text=black, transform shape, shape=rectangle, minimum size=0.5cm, thick]
       \tikzstyle{line} = [draw, -latex']
       \node at (0,0) [state, shape=circle]                      (zeta)           {$\bzeta$};
       \node [state, shape=circle, below=0.8 cm of zeta]          (x)         {$\x$};
    %   \path [line] (zeta) -- (x);
       \path [style=dashed, color=red, line, out=155, in=215] (x) edge (zeta);
       \path [line] (zeta) edge (x);
       \node [right=0.3 cm of zeta]  (prior)         {prior: $p_{\btheta}(\bzeta)$};
       \node [right=0.3 cm of x]  (model)         {joint: $p_{\btheta}(\x, \bzeta)$};
       \node [below=0.15 cm of prior]  (decoder)         {marginal: $p_{\btheta}(\x| \bzeta)$};
       \node [left=1. cm of decoder]  (posterior)         {approx. posterior: $q_{\bphi}(\bzeta|\x)$};
  \end{tikzpicture}
\caption{Generative models with latent variables can be represented as probabilistic graphical models that describe conditional relationships among variables. In a directed generative model, the joint probability distribution  $p_{\btheta}(\x, \bzeta)$, is decomposed as $p_{\btheta}(\x, \bzeta) = p_{\btheta}(\x|\bzeta) p_{\btheta}(\bzeta)$. The prior distribution over the latent variables  $p_{\btheta} (\bzeta )$ and the marginal (decoder) distribution $p_{\btheta} (\x|\bzeta )$ are hard-coded to explicitly define the model. The computation of the true posterior,  $p_{\btheta}(\bzeta| \x)$ is intractable. In VAEs, an approximating posterior $q_{\bphi}(\bzeta|\x)$ (decoder) is introduced to replace the true posterior.}
\label{fig:1}
\end{center}
\end{figure}
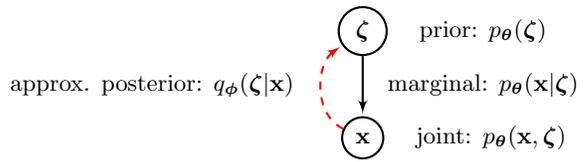

In generative modeling, the goal is to train a probabilistic model such that the model distribution $p_\btheta(\bf X)$ (where $\btheta$ are the parameters of the model) is as close as possible to the data distribution, $p_\text{data}(\bf X)$, which is unknown but assumed to exist. The ensemble ${\bf X} = \{  \x^{d}\}_{d=1}^N$ represents the training set; that is, $N$ independent and identically distributed samples coming from $p_\text{data}(\bf X)$. The preferred method to training probabilistic models is arguably maximum likelihood estimation (MLE), which means the optimal model parameters are obtained by maximizing the log-likelihood $L(\bf X, \btheta$) of the dataset with respect to the model:
\beq
L({\bf X}, \btheta)  = \sum_{\x \in \bf X} p_\text{data}(\x) \log p_{\btheta} (\x) = \Ex{[\log p_{\btheta}(\x)]}\,,
\eeq
where $\Ex[\dots]$ is the expectation over $p_\text{data}(\x)$.
Similarly to generative adversarial networks (GANs)~\cite{goodfellow_generative_2014}, VAEs~\cite{kingma2013auto} are ``directed" probabilistic models with latent variables (see Fig.~\ref{fig:1}): the model distribution, defined as the joint between the visible units $\x$ and latent units $\bzeta$, is explicitly parameterized as  the product of the ``prior" $p_{\btheta}(\bzeta)$ and ``marginal" $p_{\btheta}(\x| \bzeta)$ distributions, $p_{\btheta}(\x, \bzeta)=p_{\btheta}(\x| \bzeta) p_{\btheta}(\bzeta)$. The model prediction for the data is then obtained by marginalizing over  the  latent units:
\beq
p_{\btheta}(\x) = \int p_{\btheta}(\x|\bzeta)p_{\btheta}(\bzeta) d\bzeta\,.
\label{eq:px}
\eeq

Generative models with latent variables can potentially learn and encode  useful representations of the data in the latent space. This is an important property that can be exploited in many practical applications~\cite{fergus2009semi,liu2013graph,shi2011semi,chen2013semi} to improve other tasks such as supervised and semi-supervised  learning~\cite{kingma2014semi}. The drawback is ``intractable inference" due to the appearance of integrals such as the one in Eq.~\ref{eq:px}. Essentially, VAEs remove the necessity to evaluate such integrals by introducing a variational approximation $q_{\bphi}(\bzeta|\x)$ to the true posterior $p_{\btheta}(\bzeta|\x)$. A so-called ``reparameterization trick" is also introduced to obtain an efficient and low-variance estimate of the gradients needed for training. We will briefly review these two important elements in the next two sections.

%%%%%%%%%%%%%%%%%%%%%%%%%%%%%%%%%%%%%
\subsubsection{Variational inference}
%%%%%%%%%%%%%%%%%%%%%%%%%%%%%%%%%%%%%

Training generative models with latent variables via MLE requires the evaluation of the  intractable integral of  Eq.~\ref{eq:px} to calculate the posterior distribution $p_{\btheta}(\bzeta|\x)$. VAEs circumvent this problem by   introducing a tractable variational approximation $q_{\bphi}(\bzeta|\x)$ to the true posterior~\cite{hoffman2013stochastic}, with variational parameters $\bphi$ (see Fig.~\ref{fig:1}). VAEs are then trained by maximizing a variational lower bound $\mathcal L(\x,\btheta,\bphi)$ to the log-probabilities $ \log p_{\btheta}(\x)$:
 \bea
\mathcal L(\x, \btheta,\bphi) 
&=& \log p_{\btheta}(\x) - \E_{\bzeta\sim q_{\bphi}(\bzeta|\x)}\left[ \log \frac{q_{\bphi}(\bzeta|\x)}{p_{\btheta}(\bzeta|\x)}\right]\equiv \nn 
&\equiv& \log p_{\btheta}(\x) - D_{KL}( q_{\bphi}(\bzeta|\x) ||  p_{\btheta}(\bzeta|\x))\,,
\label{ELBOKL}
\eea
where $D_{KL}( q_{\bphi}(\bzeta|\x) ||  p_{\btheta}(\bzeta|\x))$ is the Kullback-Leibler divergence (KL divergence) between the true and approximating posteriors. Since  KL divergences are always non-negative, we have
\bea
\mathcal L(\x, \btheta,\bphi) \le \log p_{\btheta}(\x),
\eea
which immediately gives: 
\bea 
\mathcal L({\bf X}, \btheta,\bphi) & \equiv &   \Ex[ \mathcal L(\x,\btheta,\bphi)]  \nn  & \le &   \Ex[\log p_{\btheta}(\x)] \equiv L({\bf X}, \btheta)\,,
\label{eq: elbo_ge}
\eea
where $\mathcal L({\bf X}, \btheta,\bphi)$ is called the evidence lower bound (ELBO). The ELBO can be written in terms of tractable quantities: 
\bea
\mathcal L({\bf X}, \btheta,\bphi)  & = &  \Ex\Big[  \E_{\bzeta\sim q_{\bphi}(\bzeta|\x)}[\log p_{\btheta}(\x|\bzeta)]  + \nn
&- & D_{KL}( q_{\bphi}(\bzeta|\x) ||  p_{\btheta}(\bzeta))\Big]\,.
\label{eq:ELBO}
\eea 
The marginal $p_{\btheta}(\x|\bzeta)$ and approximating posterior $q_{\bphi}(\bzeta|\x)$, also called ``decoder" and ``encoder" respectively, are commonly parameterized using deep  neural networks.

%%%%%%%%%%%%%%%%%%%%%%%%%%%%%%%%%%%%%
\subsubsection{The reparameterization trick}
%%%%%%%%%%%%%%%%%%%%%%%%%%%%%%%%%%%%%
To train VAEs, we need to calculate the derivatives of the objective function (Eq.~\ref{eq:ELBO}) with respect to the generative ($\btheta$) and inference ($\bphi$) parameters.  The naive evaluation of $\pd_{\bphi}$ of terms of  the type $ \E_{\bzeta \sim q_\phi}[f(\bzeta)]$ is called REINFORCE~\cite{williams1992simple}. With the use of the identity $\pd_{\bphi} q_\phi = q_\phi \pd_{\bphi} \log q_\phi $, one has:
\bea
\pd_{\bphi} \E_{\bzeta \sim q_\phi}\left[f(\bzeta)\right] = \E_{\bzeta \sim q_\phi}\left[f(\bzeta) \pd_{\bphi} \log q_\bphi \right]\,.
\label{eq:REINFORCE}
\eea
However, the term above has high variance and requires intricate variance-reduction mechanisms to be of practical use~\cite{mnih2014neural}.

A better approach is to write the random variable $\bzeta$ as a deterministic function of the distribution parameters $\bphi$ and of an additional auxiliary random variable $\brho$. The latter is given by a probability distribution $p(\brho)$ 
that does not depend on $\bphi$. This reparameterization $\bzeta(\bphi , \brho)$ is appropriately chosen so that one can write $\E_{\bzeta \sim q_\phi}[f(\bzeta)] = \E_{\brho \sim p(\brho)}[f( \bzeta(\bphi, \brho))]$. Therefore, we can move the derivative inside the expectation with no difficulties:
\bea
\pd_{\bphi} \E_{\bzeta \sim q_\phi}\left[f(\bzeta)\right] = \E_{\brho \sim p(\brho)}\left[ \pd_{\bphi} f( \bzeta(\bphi, \brho)) \right].
\label{eq:reparam_preamble}
\eea
This is called the \emph{reparameterization  trick}~\cite{kingma2013auto} and its efficient implementation is responsible for the recent success and proliferation of VAE. 
%%%%%%%%%%%%%%%%%%%%%%%%%%%%%%%%%%%%%
\subsection{VAE with discrete latent variables}
\label{sec:DVAE}
%%%%%%%%%%%%%%%%%%%%%%%%%%%%%%%%%%%%%

The application of the reparameterization trick as in Eq.~\ref{eq:reparam_preamble} requires that $ f( \bzeta(\bphi, \brho))$ be differentiable, so the latent variables $\bzeta$ are continuous. However, discrete latent units can be indispensable to represent the right distributions, such as in attention models, language modeling, and reinforcement learning~\cite{jang2016categorical,kingma2014semi,maaloe2017semi}. For example, a latent space composed of discrete variables can learn to disentangle content and style information of images in an unsupervised fashion~\cite{makhzani2017pixelgan}.  Several methods have thus been developed to circumvent the non-differentiability of discrete latent units~\cite{paisley2012variational, mnih2014neural, gu2015muprop,bengio2013estimating}. In the context of VAE, the reparameterization trick has been extended to discrete variables by either relaxation of discrete variables into continuous variables~\cite{jang2016categorical, maddison2016concrete,khoshaman2018gumbolt} or by introducing smoothing functions~\cite{rolfe2016discrete}. In Ref.~\cite{khoshaman2018quantum}, QVAE was introduced based on the implementation of Ref.~\cite{rolfe2016discrete}. In this work, we follow the implementation of Ref.~\cite{khoshaman2018gumbolt}, which gives biased estimates but provides a much simpler and flexible implementation.

To set up a notation that we keep throughout the paper, we now assume the prior distribution is defined on a set of discrete variables  $\z \sim p_{\btheta}(\z)$, with $\z \in \{0, 1 \}^L$. Given a discrete variable $z$ with mean $q$ and logit $l=\sigma^{-1}(q) = \log(q) - \log(1-q)$ (where $\sigma=1/[1+\exp(-l)]$ is the sigmoid function), a non-differentiable implementation of Eq.~\ref{eq:reparam_preamble} for discrete variables can be obtained as
\bea
z = \Theta[\rho  - (1 - q) ] =  \Theta[\sigma^{-1}(\rho)  + l )]\,,
\eea
where  $\Theta$ is the Heaviside function and the random variable $\rho \in [0,1]$ is distributed according to a uniform distribution $\mathcal{U}$. In the second equality, we have used the fact that the inverse sigmoid function is monotonic.

A continuous smoothing (also known as the Gumbel trick~\cite{maddison2016concrete}), is performed by replacing the Heaviside function with the sigmoid function:
\bea
z =  \Theta[\sigma^{-1}(\rho)  + l ] \leadsto \zeta = \sigma\left(\frac{\sigma^{-1}(\rho)  + l }{\tau}  \right)\,,
\label{eq:smooth}
\eea
where $\tau$ is a temperature parameter introduced to control the smoothing. Typically, $\tau$ is annealed from large to low values during training. For large values of $\tau$, the bias introduced by substituting $\z$ with $\bzeta$ everywhere in the loss function is large, but the gradients propagating through $\bzeta$ are also large, facilitating training. Conversely, for low values of $\tau$ the bias is reduced but gradients vanish and training stops. Evaluation of trained models is done in the limit $\tau \rightarrow 0$, where $\bzeta \rightarrow \z$.

Throughout this paper, we will use BMs to provide powerful and expressive prior distributions defined on discrete variables:
\bea \label{eq:BD}
p_\btheta(\z) & \equiv & {e^{-\Ham_\btheta(\z)}}/Z_\btheta\,, \quad Z_\btheta \equiv  \sum_{\z} e^{-\Ham_\btheta(\z)}   \,,   \nn
\Ham_\btheta(\z) & \equiv &   \sum_l \sigma^z_l b_l  +  \sum_{l< m}  W_{lm} \sigma^z_l \sigma^z_m\,.
\eea
To train a VAE with BM prior, following the prescription of the previous section, we formally replace $p_\btheta(\z) \leadsto p_\btheta(\bzeta)$. As usual, the gradients of the log-probability is given by the difference between a positive and negative phase:
\beq 
\pd \log  p_\btheta(\bzeta_\bphi) = -  \pd  \Ham_\btheta(\bzeta_\bphi)  + \E_{\bar \z \sim p_\theta}[{\pd \Ham_\btheta(\bar \z)}]\,.
\label{eq:GT10}
\eeq
In the equation above, we have highlighted the fact that the smoothed latent samples $\bzeta_\bphi$ depend on the variational parameters $\bphi$. The model samples $\bar \z$, however, remain discrete variables sampled from the BM, and are thus not smoothed during training~\cite{khoshaman2018gumbolt}.
%%%%%%%%%%%%%%%%%%%%%%%%%%%%%%%%%%%%%
\subsection{Hybridization with quantum prior}
\label{sec:RBM-VAE}
%%%%%%%%%%%%%%%%%%%%%%%%%%%%%%%%%%%%%

Once we have a framework to train VAEs with discrete latent variables, we can consider quantum-classical hybrid VAEs in which the generative process $\z \sim p_\btheta(\z)$ is realized by measuring the computational basis on a given quantum state $\rho_\btheta$. Such quantum states can be realized by a quantum circuit or via a quantum annealing process controlled by a set of parameters $\btheta$ we wish to adjust during training of the model.

As introduced in Ref.~\cite{khoshaman2018quantum}, a QVAE can be obtained by assuming the quantum state  $\rho_\btheta$ is a thermal state of a transverse field Ising model; \emph{i.e.}, a QBM~\cite{Amin:2016qd}. The prior  $p_\btheta(\z)$ distribution is then given by:
\bea \label{eq:QBD}
p_\btheta(\z) & \equiv & \tr[\Lambda_\z {e^{-\Ham_\btheta}}]/Z_\btheta\,, \quad Z_\btheta \equiv  \tr [e^{-\Ham_\btheta}]   \,,   \nn
\Ham_\btheta& =&  \sum_l \sigma^x_l \Gamma_l + \sum_l \sigma^z_l b_l  +  \sum_{l< m}  W_{lm} \sigma^z_l \sigma^z_m,
\eea
where ${\bf \Gamma}, {\bf h}, {\bf W}  \in \{\btheta \}$,  $\Lambda_\z \equiv | \z \rangle \langle \z |$ is the projector on the classical state $\z$, and $\sigma_l^{x,z}$ are Pauli operators. Unlike for a classical BM, the direct evaluation of the gradients of the term above is intractable. As discussed in Ref.~\cite{Amin:2016qd}, a possible workaround is to perform the following substitution:
\bea
p_\btheta(\z) = \tr[\Lambda_\z {e^{-\Ham_\btheta}}]/Z_\btheta \rightarrow \tilde p_\btheta(\z) = e^{-\Ham_\btheta(\z)}/Z_\btheta
\eea
in the ELBO $\mathcal L$ to obtain the so-called quantum ELBO (Q-ELBO) $\tilde{\mathcal L}$. As a consequence of the Golden-Thompson inequality $\tr[ e^{A} e^{B}] \ge \tr [e^{A + B}]$, one has:
\bea
p_\btheta(\z) \ge \tilde p_\btheta(\z) \quad \Rightarrow \quad \mathcal L \ge \tilde{\mathcal L}\,.
\eea
The Q-ELBO $ \tilde{\mathcal L}$ is thus a lower bound to the ELBO with tractable gradients that can be used during training. The derivatives of the log-probabilities $\log \tilde p_\btheta(\z)$  can be estimated via sampling from the QBM~\cite{Amin:2016qd}:
\beq 
\pd \log \tilde p_\btheta(\z) = -  \pd  \Ham_\btheta(\z)  + \E_{\bar \z \sim p_\theta}[{\pd \Ham_\btheta(\bar \z)}]\,,
\label{eq:GT2}
\eeq
where $\bar \z$ are the model samples distributed according to the quantum Boltzmann distribution. The use of the Q-ELBO and its gradients precludes the training of the transverse field $\bf \Gamma$~\cite{Amin:2016qd}, which is treated as a constant (hyper-parameter) throughout the training.  Training via the Q-ELBO is performed as in the BM case, by smoothing $\z \leadsto \bzeta$.

%%%%%%%%%%%%%%%%%%%%%%%%%%%%%%%%%%%%%%
\section{Sampling with Quantum Annealers}
\label{sec:QA}
%%%%%%%%%%%%%%%%%%%%%%%%%%%%%%%%%%%%%%
Currently manufactured quantum annealers physically implement a transverse-field Ising model:
\bea
\Ham(s)  = A(s) \sum_l \sigma^x_l  + B(s) \left[\sum_l \sigma^z_l h_l  +  \sum_{l< m}  J_{lm} \sigma^z_l \sigma^z_m\right]\,,\nonumber
\eea
where $s\in [0, 1]$ is a control parameter, and $A(s)$ and $B(s)$ are respectively decreasing and increasing monotonic functions of the parameter $s$ with $A(0) \gg B(0)$ and $A(1) \ll B(1)$.
%A typical quantum annealing routine prescribes performing a slow anneal $s=t/t_a$, where $t$ is time and $t_a$ is the total annealing time. Thanks to the adiabatic theorem, if the system above is initialized in the ground state of $\Ham(0)$, it stays in the instantaneous ground state.  The implementation of the system above allows the solution of a classical quadratic optimization problem whose solution is encoded in the ground state of $\Ham(1)$, provided the annealing time $t_a$ is sufficiently large.
Quantum annealers operate immersed in a thermal environment. %Thermal fluctuations prevents the system to stay in the instantaneous ground state and will have a detrimental effect to the performance of quantum annealers as optimizers.
There is theoretical and numerical evidence~\cite{amin2015searching, marshall2017thermalization,marshall2019power, raymond2016global} that when the anneal is performed sufficiently slowly the system above is in thermal equilibrium with the environment. This property can be exploited to turn quantum annealers into programmable Boltzmann samplers. Thermal relaxation rates are controlled by the intensity of the transverse field $A(s)$. At the beginning of the anneal, relaxation times are small, and the system proceeds through a sequence of thermal states. As the anneal proceeds, relaxation times grow larger and eventually the state of the system freezes at the point $s^{*}$ where relaxation times roughly become larger than the annealing time $t_a$.

With the above picture in mind, we can use quantum annealers to sample from the QBM defined in Eq.~\ref{eq:QBD} with:
\bea
 b_{l} &=& \beta^{*}_{eff} h_{l}, \quad W_{lm} = \beta^{*}_{eff} J_{lm},  \quad \Gamma_{l} = \beta^{*}_{eff} \Gamma^{*},\nn
\beta_{eff} & \equiv & B(s^{*}) / \beta_{phys}, \quad \Gamma^{*} \equiv A(s^{*}) /B(s^{*})\,.
\label{eq:freez}
\eea
Advanced control techniques for the anneal schedule (such as pauses and fast ramps present in the latest generation of D-Wave quantum annealers) allow in principle to control the freezing point $s^{*}$.

Note that knowledge of the effective transverse field $\Gamma^{*}$ is unnecessary: QBMs are trained via the Q-ELBO, in which the transverse field does not appear explicitly, but only implicitly in sampling from the model. In the following we assume that for the models and datasets under consideration freezing happens late in the anneal. This means we effectively sample from a QBM that is very close to a classical BM. More specifically, we use quantum annealers to quantum-assist training of VAEs with classical latent-space BMs.

%%%%%%%%%%%%%%%%%%%%%%%%%%%%%%%%%%%%%
\section{Training VAE with quantum annealers}
\label{sec:HW}
%%%%%%%%%%%%%%%%%%%%%%%%%%%%%%%%%%%%%
We have implemented a convolutional VAE  whose prior is implemented by a BM. To improve the performance of the model, we use several techniques such as learning-rate and KL-term annealing, importance-weight annealing, convolution gating, and  batch normalization. We give a detailed description of the model in Appendix~\ref{sec:CONVVAE}. In this section, we restrict ourselves to a Chimera structured restricted BM (RBM) with $288$ latent units (a six-by-six patch of Chimera cells) and present our results with models trained end-to-end by using samples drawn with D-Wave 2000Q quantum annealers on the common handwritten digit dataset  MNIST~\cite{lecun1998mnist}. Samples used to estimate the negative phase (second term of Eq.~\ref{eq:GT2}) are obtained following the prescription given in Appendix~\ref{sec:exp}. 

\begin{figure}[t]
\begin{center}
%\subfigure[\, Chimera]{
\includegraphics[width=0.8\columnwidth]{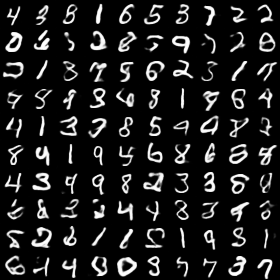}
%
%\subfigure[\, Bernoulli]{\includegraphics[width=0.4\columnwidth]{Bernoulli}\label{fig:3b}}
\caption{Images generated by sampling latent configurations with a quantum annealer that are subsequently transformed by a classical deconvolutional decoder. The classical networks and the quantum annealer weights have been trained end-to-end for 2000 epochs on the MNIST dataset  using the quantum annealer as a sampler for estimating the gradients of the annealing parameters.}
\label{fig:2}
\end{center}
\end{figure}

The effective temperature $ \beta^{*}_{eff}$ must be chosen appropriately to correctly train the parameters of the inference network $q_\bphi$ (see appendix~\ref{sec:betaest} for a more detailed discussion). The parameter  $ \beta^{*}_{eff}$ can be considered as a multiplicative correction for the learning rate of the prior parameters $b, W \in \btheta$. However, this observation is not true for the inference parameters $\bphi$, whose gradients also propagate through the first term in Eq.~\ref{eq:GT2} via $\bzeta(\bphi, \brho)$. Due to our simple forward-anneal schedule, we expect the value of $ \beta^{*}_{eff}$ to change during training. To account for this effect, in this work we employ a real-time $ \beta^{*}_{eff}$ estimation as explained in Appendix~\ref{sec:betaest}. In future works, we expect to be able to train at a fixed-temperature with appropriate pause-and-ramp annealing schedules such as those available with D-Wave quantum annealers~\cite{harris2018phase, king2018observation}.

Training is performed jointly on the parameters of the classical networks and on the parameters of the quantum device. The gradients of the latter parameters require estimation of the negative phase (a thermal expectation of the energy) in Eq.~\ref{eq:GT2}. At each gradient update, such expectations are computed using samples from the quantum annealer only, and do not involve any classical Gibbs sampling such as persistent contrastive divergence, or any classical post-processing of the samples obtained by the annealer. We typically trained our models for $2000$ epochs and a batch size of $1000$. Figure~\ref{fig:2} shows a set of images generated by a VAE trained end-to-end using a D-Wave 2000Q system. The set of images, obtained by generating latent samples $\z$ with the quantum annealer and subsequently decoded as $\x \sim p_{\btheta}(\x| \z)$, shows a good amount of global consistency and consistent statistical variety.

%%%%%%%%%%%%%%%%%%%%%%%%%%%%%%%%%%%%%
\subsection{Validation of training}
\label{sec:VALID}
%%%%%%%%%%%%%%%%%%%%%%%%%%%%%%%%%%%%%
\begin{figure}[b]
\begin{center}
\subfigure[\, Chimera]{\includegraphics[width=0.45\columnwidth]{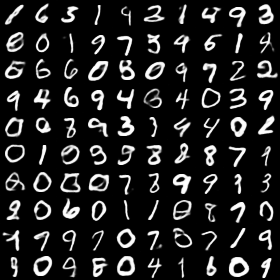}\label{fig:3a}}
\hspace{0.1cm}
\subfigure[\, Bernoulli]{\includegraphics[width=0.45\columnwidth]{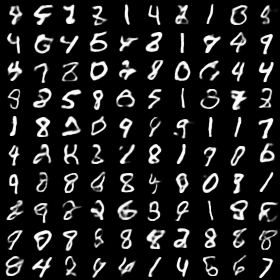}\label{fig:3b}}
\caption{Visual validation of training with Chimera connectivities is not conclusive.}
\label{fig:3}
\end{center}
\end{figure}

In this section we give more evidence that we have successfully exploited the Chimera-structured RBM prior in the latent space of our convolutional VAE. Validating the training of quantum-classical hybrid generative models can be nontrivial and must be assessed carefully, especially when training uses a large amount of classical processing. We trained the deep networks of our model using GTX 1080 Ti GPUs, and the quantum annealer is called only to estimate the negative phase. A principled validation strategy is  thus to compare a model trained with quantum assistance to a fully classical baseline for which the quantum hardware is not required. In our case, a convenient baseline is a model that has the same classical networks but whose prior is trivial. As a trivial prior, we choose  a set of independent Bernoulli variables, which is equivalent to an RBM with vanishing weights between latent units. For simplicity, in the following we refer to such prior as RBM with Bernoulli prior.
%While generative models with latent variables are also powerful tools to perform supervised and semisupervised tasks, in this work we focus on training VAE as purely generative models. The ultimate goal of a generative model is thus to generate samples that are plausible representations of the training (or more precisely the test) set. This is achieved maximizing the log probability that the test is generated by the model (maximum likelihood estimation). 

For image processing, comparison between different generative models could be done qualitatively by visually inspecting the generated samples. In our research however, visual comparison is inconclusive~\cite{theis2015note}. In Fig.~\ref{fig:3}, for example, we compare samples generated by a trained model with Chimera (Fig.~\ref{fig:3a}) and Bernoulli (Fig.~\ref{fig:3b}) priors. Both models have been trained by evaluating the negative phase with a D-Wave 2000Q system. The images are also generated using samples coming from the annealers. Given our specific implementation, it is difficult to discern an improvement of visual quality when using a Chimera-structured RBM rather than a Bernoulli prior. To overcome this difficulty, we evaluate quantitatively~\footnote{The existence of a well-defined loss function that allows quantitative validation is an important advantage of VAE, compared to other generative models such as Generative Adversarial Networks} the generative performance of VAEs by computing the ELBO defined in Eq.~\ref{ELBOKL} or by estimating the log-likelihood  via an importance sampling technique as described in Ref.~\cite{burda2015importance}. 

These quantities are however not accessible when training with analog devices as samplers. In fact, while we assume that samples generated by quantum annealers are distributed according to the required Boltzmann distribution, we must treat quantum annealers as black-box samplers during testing and validation. In other words, the log-probabilities for the quantum generative process $\log p^{\rm DW}_{\btheta}(\z)$ must be assumed to be unknown. Therefore, we validate results by replacing the unknown hardware log-probabilities with those of an auxiliary 
RBM whose weights are given by the relations in Eq.~\ref{eq:freez}~\footnote{The computation of the log-probabilities requires the estimation of the partition function, which is done using annealed population sampling as in Ref.~\cite{khoshaman2018quantum}}:
\bea
\log p^{\rm DW}_{h, J}(\z)  \leadsto \log p^{\rm RBM}_{b, W}(\z)\,. 
\eea
This approach can be more rigorously interpreted as validating the fully classical model in which we replace the quantum annealer by the auxiliary RBM defined by the relations above.
 
 In Tab.~\ref{tab:auxval} we report the LL of the auxiliary VAE with 288 latent unit on a Chimera connectivity trained with a D-Wave 2000Q quantum annealer. We compare it to the same model trained end-to-end with population annealing~\cite{khoshaman2018quantum}. We also compare each model with its respective Bernoulli baseline. We have reported the mean and the standard error over 5 independent training runs.
\begin{table}[t]
\begin{center}
\begin{tabular}{|lll|}
\hline
\multicolumn{3}{|c|}{MNIST (dynamic binarization) {\bf LL}  }\\
 \hline
Sampler & Chimera  & Bernoulli\\
\hline
 DW2000Q \hspace{1cm}  & $ -82.8\pm 0.2$  &$-83.7\pm 0.2$ \\
 PA \hspace{1cm}  & $ -82.8\pm 0.1$  &$-84.2\pm 0.05$ \\
\hline
\end{tabular}
\end{center}
\caption{Log-likelihood of convolutional VAEs trained with samples coming from either D-Wave 2000Q or PA. All models share the same encoding and decoding networks, but are trained independently for 2000 epochs on the MNIST dataset.}
\label{tab:auxval}
\end{table}%
Training with a Chimera-structured RBM improves significantly the log-likelihood over the Bernoulli baseline. Moreover, the models trained with quantum annealers achieved the same log-likelihood as the models trained with PA. 
% It is not surprising that a model with Chimera structured prior improves over its Bernoulli baseline, since it is a model with strictly more capacity and trainable weights. We indeed use this expected improvement to validate we successfully trained an Chimera structured RBM using quantum annealers as samplers. 
Notice that each model and its baseline employed exactly the same amount of classical computational resources. Models trained with structured RBMs achieve better performance by requiring thermal sampling, a computational task that can be offloaded to a quantum annealer.

\begin{figure}[t]
\begin{center}
\includegraphics[width=0.45\columnwidth]{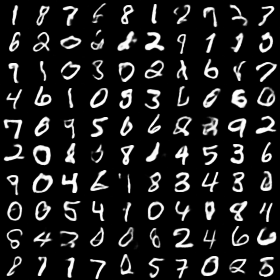}
\hspace{0.1cm}
\includegraphics[width=0.45\columnwidth]{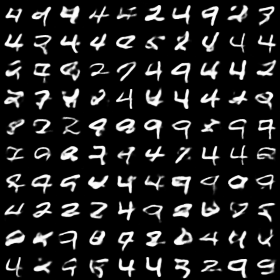}
\caption{Left panel: Samples generated by D-Wave 2000Q using a fully trained model. Right panel:  Samples generated by D-Wave 2000Q after setting the couplings of the annealer to zero.}
\label{fig:4}
\end{center}
\end{figure}

%\subsubsection{Validation of hardware samples for generative purposes}
In general we would like to use quantum annealers to sample from the trained generative model. As explained above, treating quantum annealers as black-boxes means we cannot quantitatively evaluate such a model. However, we argue that the log-likelihood of such a VAE is likely very close to that of the auxiliary VAE. After all, the training assumes the hardware samples are distributed according to a Boltzmann distribution of the auxiliary RBM, and in the previous section we have shown that this assumption is accurate enough to correctly train the auxiliary RBM. We can confirm this visually in Fig.~\ref{fig:4}. On the left panel we show a set of digits generated by sampling from a D-Wave 2000Q. On the right panel we use the same trained model but sample from a D-Wave 2000Q after setting its weights to zero. We see that while the annealer still generates plausible digits, it does not generate a number of digits with the correct statistics (in the right panel of Fig.~\ref{fig:4}, digits 9 and 4 seem to dominate the scene). This gives evidence that D-Wave 2000Q quantum annealers sampled consistently, such that the classical networks were able to correctly learn the correlations between latent units existing due to the RBM with non-vanishing weights.

%We can also consider a quantitative comparison by evaluating the entropy of the auxiliary model and the cross-entropy of the same mode the hardware model:
%\bea
%H(p^{RBM}) = \E_{\x \sim p^{RBM}(\x)}[ \log p^{RBM}(\x)] \nonumber \\
%H(p^{HW}, p^{RBM}) = \E_{\x \sim p^{HW}(\x)} [\log p^{RBM}(\x)]\,.
%\eea
%We would like the two quantities to be close. We compare their value to reference cross-entropies in which we sample from an RBM without weights and from a random uniform prior distribution.

%%%%%%%%%%%%%%%%%%%%%%%%%%%%%%%%%%%%%
\section{A path towards quantum advantage with VAE}
\label{sec:path}
%%%%%%%%%%%%%%%%%%%%%%%%%%%%%%%%%%%%%

In the previous section we have shown that it is possible to use quantum annealers as Boltzmann samplers to train RBM-structured priors placed in the latent space of deep convolutional VAE. In the experiments presented, we have settled on relatively small, Chimera-structured RBMs with 288 latent units. We found that larger RBMs did not appreciably improve performance of the overall VAE when training on the MNIST dataset. We will explain why this is the case in this section. Sampling from RBMs with a few hundred units can still be done classically with relative ease, and the natural question that arises is whether we can obtain a quantum advantage in this hybrid setup.

A necessary condition to obtaining quantum advantage with our QCH setup is to engineer  models that can exploit very large RBMs with complex, multimodal distributions. Sampling from RBMs with complex landscapes is slow and inefficient for classical approaches such as Gibbs sampling, PCD, and PA. Another important requirement is to be able to reliably sample from thermal states using quantum annealers with sufficiently low control errors.

In the next sections, we give evidence of the existence of a natural path towards obtaining quantum advantage by applying quantum annealing to generative modeling within the proposed VAE framework.

%%%%%%%%%%%%%%%%%%%%%%%%%%%%%%%%%%%%%
\subsection{Exploit large latent-space RBMs}
\label{sec:RBM}
%%%%%%%%%%%%%%%%%%%%%%%%%%%%%%%%%%%%%

Exploiting a larger number of latent units to improve the generative performance of a VAE is a popular and active research area. One known obstacle in achieving this is the loss function used for training. We rewrite the ELBO here for convenience by highlighting its two terms:
\bea
\mathcal L(\x, \z,\bphi)  & = &    \underbrace{\E_{\z\sim q_{\bphi}(\z|\x)}[\log p_{\btheta}(\x|\z)]}_{\rm autoencoding \, term}  + \nn
&- & \underbrace{D_{KL}( q_{\bphi}(\z|\x) ||  p_{\z}(\z))}_{\rm KL-regularization}\,.
\label{eq:ELBO_split}
\eea 
The first term is sometimes called the ``autoencoding" term and can be thought of as a reconstruction error: an encoded latent configuration $\zeta$ is first sampled from the encoder $q_{\bphi}(\bzeta|\x)$ and is subsequently decoded by $p_{\btheta}(\x|\bzeta)$. When both approximating posterior and marginal distributions are factorized distributions, this term can be also interpreted as a reconstruction error. Maximizing the ELBO results in maximizing this term, which tends to maximize the number of latent units used to prevent information loss during the encoding-decoding steps. The second term, the KL divergence between the approximating posterior and the prior, has the effect of a regularization term and it is sometimes also called KL regularization. Maximizing the ELBO results in minimizing the KL term, which pushes the approximating posterior close to the prior. This also means the approximating posterior depends less sharply on the inputs $\x$. In the case of factorized distributions, this usually means some latent units are conditionally independent from the input (``inactive units"): $z_{inact} \sim q_\btheta(z_{inact}|\x) = q_\btheta(z_{inact})$.

\begin{figure}[t]
\begin{center}
\includegraphics[width=1\columnwidth]{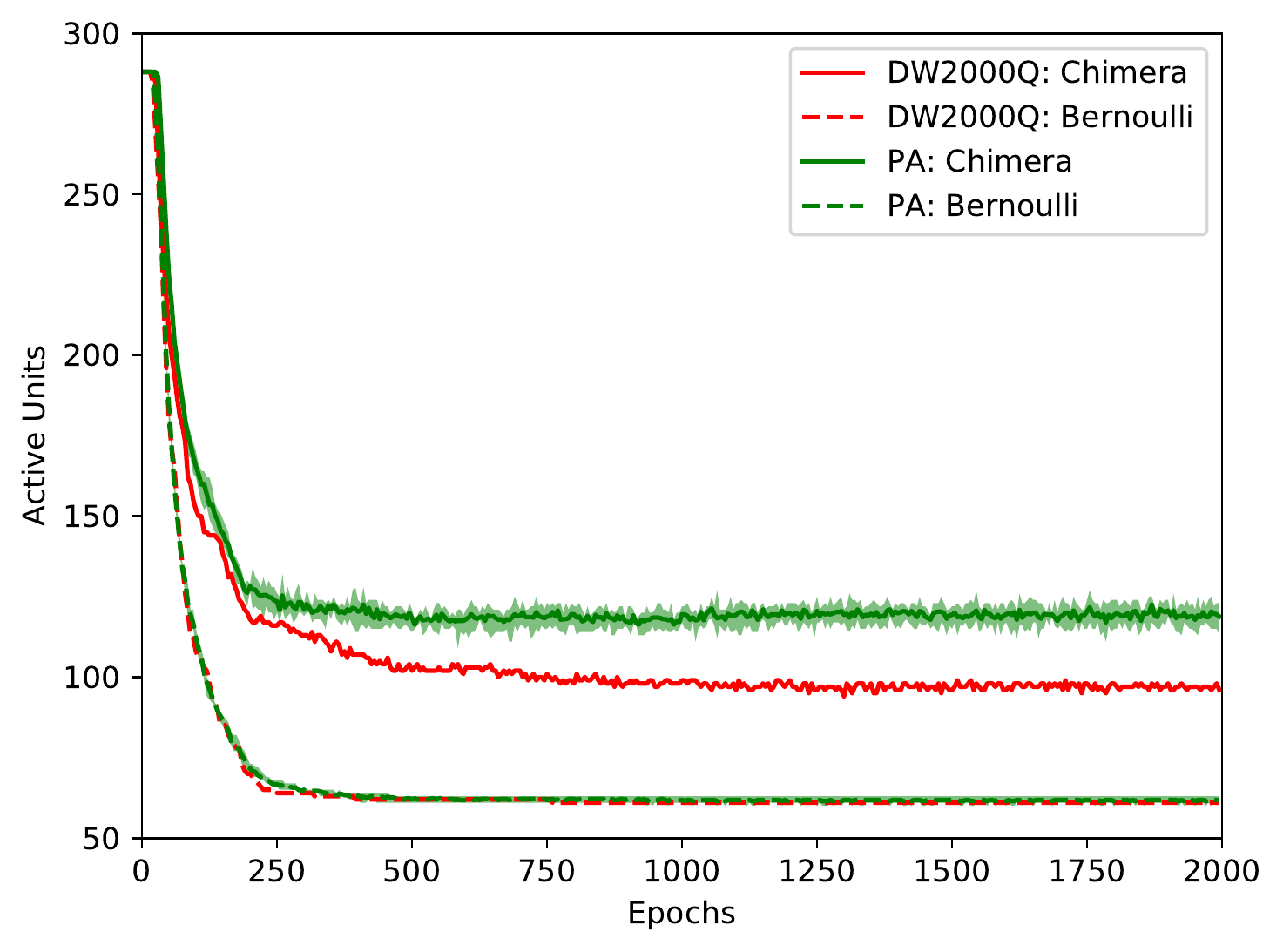}
\caption{Comparison of active units during training between a D-Wave 2000Q and population annealing on both Chimera and Bernoulli priors.}
\label{fig:5}
\end{center}
\end{figure}

The balance between the autoencoding and KL terms means using VAE can be efficient at lossy data compression by using the right number of latent units. However, the tension between KL and autoencoding terms results in an optimization challenge: during training the model is usually stuck in a local minimum with a suboptimal number of latent units. The main takeaway, for our purposes, is that the number of latent units effectively used is highly dependent on the model, the optimization technique, and the training set. To exploit a larger RBM, one thus has to work on all these elements. In Fig.~\ref{fig:5} we show the number of active units during a training run of 2000 epochs when the models are trained with either PA or D-Wave 2000Q, with either a Chimera-structured RBM or a Bernoulli prior. Lines (bold or dashed) are the means over 5 independent runs while light-color areas delimit the smallest and largest values among the 5 runs. To identify whether a latent unit is active or not, we compute the variance $\sigma$ of the value of each unit $z$ over the test set and we set a threshold of $\sigma > 0.01$ as definition of an active unit.

Figure~\ref{fig:5} shows several key points that we will expand upon in the next sections. First, it shows the use of a KL annealing technique: the KL term is turned off at the beginning of the training and it is slowly (linearly) turned on within 200 epochs. The figure clearly shows the effect of the KL term in shutting down a large number of active units. Since the number of active units plateaus around a number much lower than 288, using a larger RBM usually does not improve performance of our implementation on MNIST. It also shows that connectivity of the RBM plays a major role in determining the number of active units, which is much higher with a Chimera-structured RBM. Notice also that in the Bernoulli case, both samplers (PA and D-Wave 2000Q) train a model that uses a very similar number of latent units. However, when sampling is nontrivial (as in a Chimera-structured RBM) the model trained with the quantum annealer uses a number of units larger than the Bernoulli case, but smaller than the model trained with PA. This is a manifestation (which we will discuss later) of biased sampling with the quantum annealer: sampling quality is good enough to train the model (indeed we obtained a log-likelihood as good as that of the model trained with PA) but exploits a smaller number of latent units.

\begin{figure*}[t]
\begin{center}
% \subfigure[\, Log-likelihood obtained with our convolutional VAE ]{
\includegraphics[width=1\columnwidth]{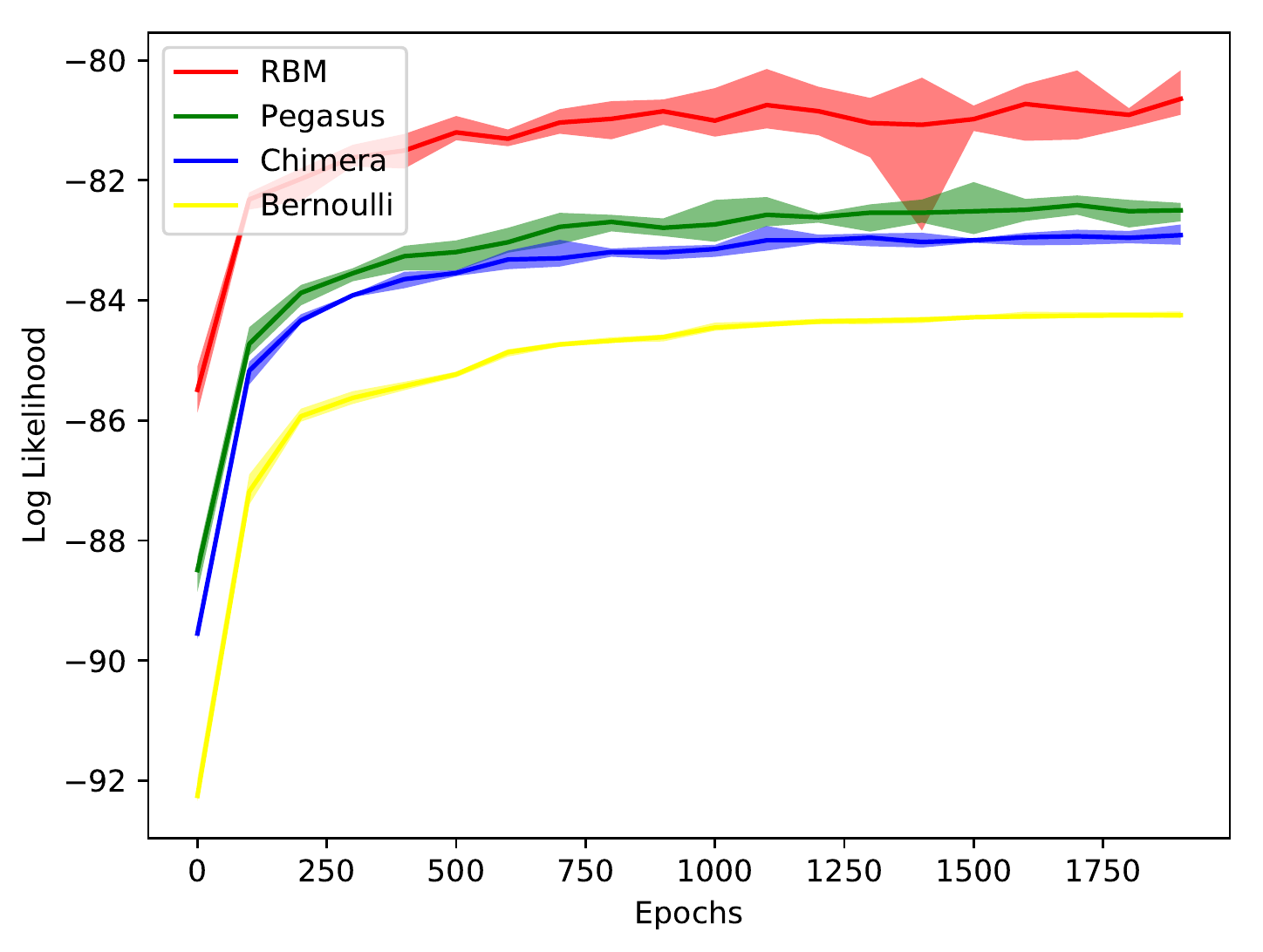}
% \label{fig:7a}}
% \subfigure[\, b]{
\includegraphics[width=1\columnwidth]{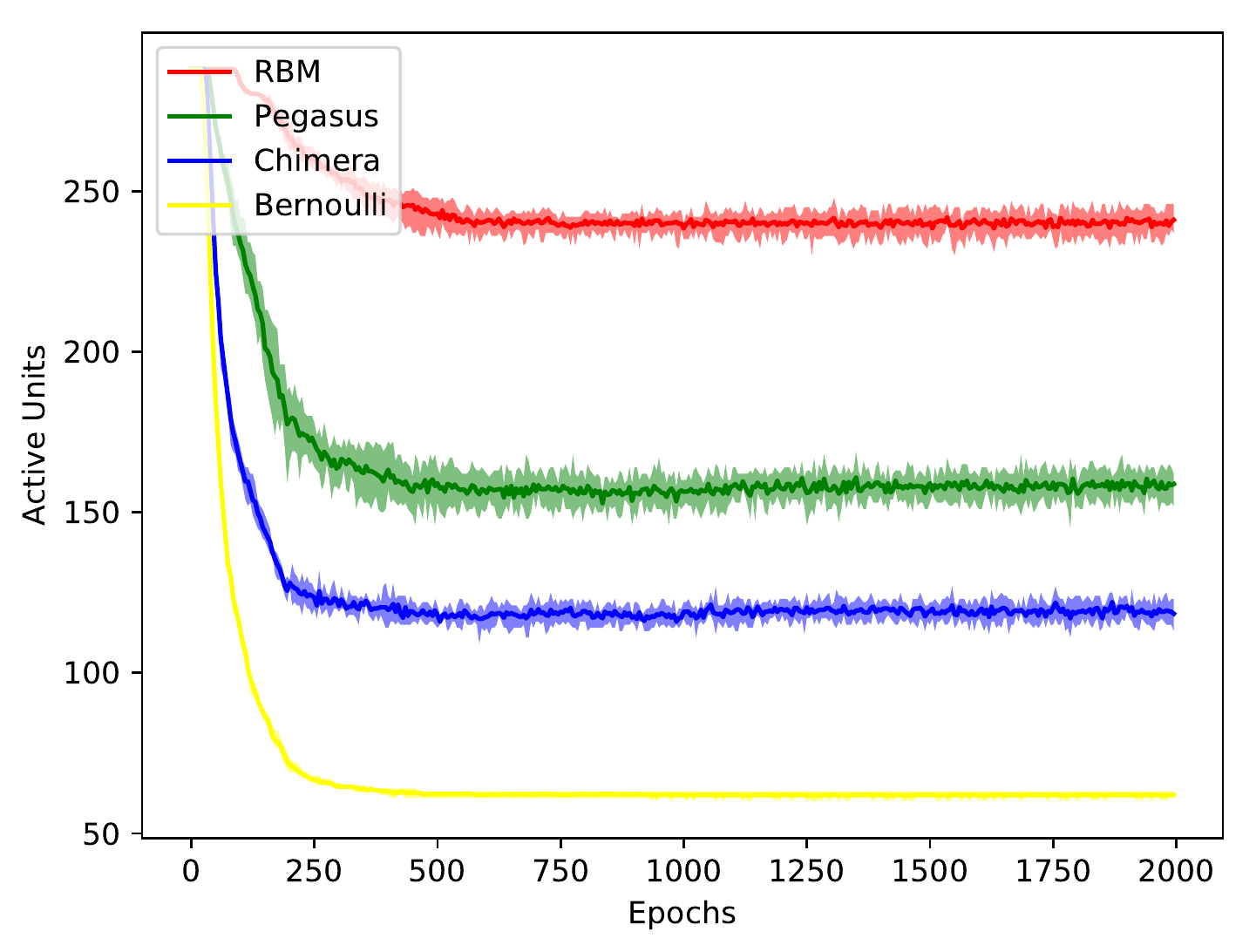}
% \label{fig:7b}}\\
\caption{Comparison between VAE using different latent-space RBM connectivities and same convolutional networks. Both log-likelihood (left panel) and latent-space utilization (right panel) improve with more densely connected RBM. Performance on Pegasus graph increases despite the development of an architecture fairly optimized for Chimera graphs.}
\label{fig:6}
\end{center}
\end{figure*}

In the next three sections we discuss three important elements that would allow to build quantum-classical hybrid VAEs that can effectively exploit large latent-space RBMs. This is a necessary condition to search for quantum advantage in these models: speed-up and scale-up sampling from large RBMs using quantum annealers rather than inefficient classical sampling techniques.

%%%%%%%%%%%%%%%%%%%%%%%%%%%%%%%%%%%%%
\subsubsection{Denser connectivities}
\label{sec:CONN}
%%%%%%%%%%%%%%%%%%%%%%%%%%%%%%%%%%%%%

Connectivity of the RBM in the latent space plays an important role in determining both performance of the generative model and number of active latent units. While these two elements are not directly related, we typically observe a correlation between them. In this section, we investigate in more detail the effects of implementing denser connectivities by performing numerical experiments in four different cases: Bernoulli prior and Chimera, Pegasus, and fully connected RBM. Together with Bernoulli and RBM, we pick the connectivities  of currently available D-Wave 2000Q quantum annealers and D-Wave's next-generation Pegasus architecture~\cite{boothby2019next} (see also Appendix~\ref{sec:CHI-PEG}).

In Fig.~\ref{fig:6} we compare the log-likelihood and the number of active units along training runs obtained using the same classical networks but using a different connectivity for the latent-space RBM prior. At each step during the training run, we employed roughly the same amount of classical resources for back-propagation. Unsurprisingly, using a more capable (dense) RBM results in better generative performance (log-likelihood) of the model (left panel)). It also results in using a larger number of latent units (right panel). Notice how Bernoulli and Chimera priors effectively use a number of latent units well below 150, while Pegasus and fully connected use well above 150. Because of this, we could not improve generative performance of Bernoulli and Chimera models by just using larger graphs, at least when training on MNIST. On the other hand, using larger Pegasus and fully connected RBMs would likely have improved the log-likelihood of the model. In Fig.~\ref{fig:6} we show results on the same number of latent units (288) for proper comparison.

Working with VAEs allows us to easily take advantage of new connectivities without having to implement a new convolutional VAE. In fact, our model has been fairly optimized to improve performance on Chimera graphs (see the next section). Despite this architecture-specific optimization, just using the denser Pegasus graphs improve performance of the model. Moreover, the flexibility of the VAE hybrid approach allows us to easily adapt the implementation to the slightly different working graphs of different processors with different  active/inactive qubits. By using the same implementation, during training the model naturally learns to deactivate latent units corresponding to uncalibrated qubits. We have indeed seamlessly used the same model to train on different D-Wave 2000Q processors, as well as using different groups of qubits within the same processor. In no case was a hard-coded connectivity (which would change for each processor) necessary.

The results of Fig.~\ref{fig:6} show that developing quantum annealers with denser connectivities (such as Pegasus) naturally leads to exploiting larger latent-space RBMs, possibly getting us closer to a regime where quantum advantage is possible.

%%%%%%%%%%%%%%%%%%%%%%%%%%%%%%%%%%%%%
\subsubsection{Hardware-specific optimization of classical networks}
\label{sec:OPT}
%%%%%%%%%%%%%%%%%%%%%%%%%%%%%%%%%%%%%

\begin{figure*}[t]
\begin{center}
\subfigure[\, Conditional posterior]{
  \begin{tikzpicture}[->,>=stealth',shorten >=1pt,auto,node distance=1.0cm, thick, scale=0.85]
       \tikzstyle{every state}=[fill=white,draw=black,text=black, transform shape, shape=rectangle, minimum size=0.7cm, ultra thick]
       \tikzstyle{line} = [draw, -latex']
       \node at (0,0) [state, shape=circle]                      (x)           {$\x$};
       \node [state, right=0.4 of x]                  (conv)        {convolutional};
       \path [line] (x) -- (conv);
       \node [state, draw=orange, below=0.8 cm of conv]          (d1)         {dense};
       \path [line] (conv) -- (d1);
       \node [state, draw=orange, below=0.8 cm of d1]            (r1)         {$q(\z_1|\x)$};
       \node [state, draw=orange, shape=circle, right=0.8 cm of r1]           (z1)         {$\z_1$};
       \path [line] (d1) -- (r1);
       \path [line, dotted] (r1) -- (z1);
       \node [state, draw=teal, right=2.4 cm of d1]            (d2)         {dense};
       \node [state, draw=teal, below=0.8 cm of d2]            (r2)         {$q(\z_2|\z_1, \x)$};
       \path [line] (conv) -| (d2);
       \node [state,  draw=teal, shape=circle, right=0.8 cm of r2]          (z2)         {$\z_2$};
       \path [line] (d2) -- (r2);
       \path [line] (z1) |-  (d2);
       \path [line, dotted] (r2) --  (z2);
      \node [state, draw=white, shape=circle, below=2.6 cm of x]           (w)    {};
  \end{tikzpicture}\label{fig:7a}}
  \quad\quad
  \subfigure[\, Bipartite mapping]{
      \begin{tikzpicture}
    \foreach \xx in {0, 6, 12}
    \foreach \yy in {0, 6, 12}
    \foreach \x  in {2, 3, 5, 6}
    \foreach \y  in {2, 3, 5, 6}
    {
    \draw (0.2*\x+0.2*\xx,0.2*4+0.2*\yy) -- (0.2*4+0.2*\xx,0.2*\y+0.2*\yy);
    }
    \foreach \x in {2, 3, 5, 6, 8, 9, 11, 12, 14, 15, 17, 18}
    {
    \filldraw [white] (0.2*1,0.2*\x) circle (2pt);
    \filldraw [white] (0.2*19,0.2*\x) circle (2pt);
    \filldraw [white] (0.2*\x,0.2*1) circle (2pt);
    \filldraw [white] (0.2*\x,0.2*19) circle (2pt);
    \draw (0.2*\x,0.2) -- (0.2*\x,0.2*19);
    \draw (0.2,0.2*\x) -- (0.2*19,0.2*\x);
    }
    
    \foreach \y in {2, 3, 5, 6}
    {
    \filldraw [orange] (0.2*\y+0.2*0,0.2*4) circle (2pt);
    \filldraw [orange] (0.2*\y+0.2*0,0.2*16) circle (2pt);
    \filldraw [orange] (0.2*\y+0.2*6,0.2*10) circle (2pt);
    \filldraw [orange] (0.2*\y+0.2*12,0.2*4) circle (2pt);
    \filldraw [orange] (0.2*\y+0.2*12,0.2*16) circle (2pt);
    \filldraw [teal] (0.2*4,0.2*\y+0.2*0) circle (2pt);
    \filldraw [teal] (0.2*4,0.2*\y+0.2*12) circle (2pt);
    \filldraw [teal] (0.2*10,0.2*\y+0.2*6) circle (2pt);
    \filldraw [teal] (0.2*16,0.2*\y+0.2*0) circle (2pt);
    \filldraw [teal] (0.2*16,0.2*\y+0.2*12) circle (2pt);
    \filldraw [teal] (0.2*\y+0.2*6,0.2*4) circle (2pt);
    \filldraw [teal] (0.2*\y+0.2*0,0.2*10) circle (2pt);
    \filldraw [teal] (0.2*\y+0.2*12,0.2*10) circle (2pt);
    \filldraw [teal] (0.2*\y+0.2*6,0.2*16) circle (2pt);
    \filldraw [orange] (0.2*10,0.2*\y+0.2*0) circle (2pt);
    \filldraw [orange] (0.2*4,0.2*\y+0.2*6) circle (2pt);
    \filldraw [orange] (0.2*16,0.2*\y+0.2*6) circle (2pt);
    \filldraw [orange] (0.2*10,0.2*\y+0.2*12) circle (2pt);
    }
    \end{tikzpicture}
    \label{fig:7b}
    }
    \quad \quad
  \subfigure[\, Chains mapping]{
    \begin{tikzpicture}
    \foreach \xx in {0, 6, 12}
    \foreach \yy in {0, 6, 12}
    \foreach \x  in {2, 3, 5, 6}
    \foreach \y  in {2, 3, 5, 6}
    {
    \draw (0.2*\x+0.2*\xx,0.2*4+0.2*\yy) -- (0.2*4+0.2*\xx,0.2*\y+0.2*\yy);
    }
    \foreach \x in {2, 3, 5, 6, 8, 9, 11, 12, 14, 15, 17, 18}
    {
    \filldraw [white] (0.2*1,0.2*\x) circle (2pt);
    \filldraw [white] (0.2*19,0.2*\x) circle (2pt);
    \filldraw [white] (0.2*\x,0.2*1) circle (2pt);
    \filldraw [white] (0.2*\x,0.2*19) circle (2pt);
    \draw (0.2*\x,0.2) -- (0.2*\x,0.2*19);
    \draw (0.2,0.2*\x) -- (0.2*19,0.2*\x);
    }
    \foreach \yy in {0, 6, 12}
    \foreach \x in {4, 10, 16}
    \foreach \y in {2, 3, 5, 6}
    {
    \filldraw [teal] (0.2*\x,0.2*\y+0.2*\yy) circle (2pt);
    \filldraw [orange] (0.2*\y+0.2*\yy,0.2*\x) circle (2pt);
    }
    \end{tikzpicture}
    \label{fig:7c}
    }
\caption{Implementation of a conditional approximate posterior (left), with two possible mappings between groups.}
\end{center}
\end{figure*}
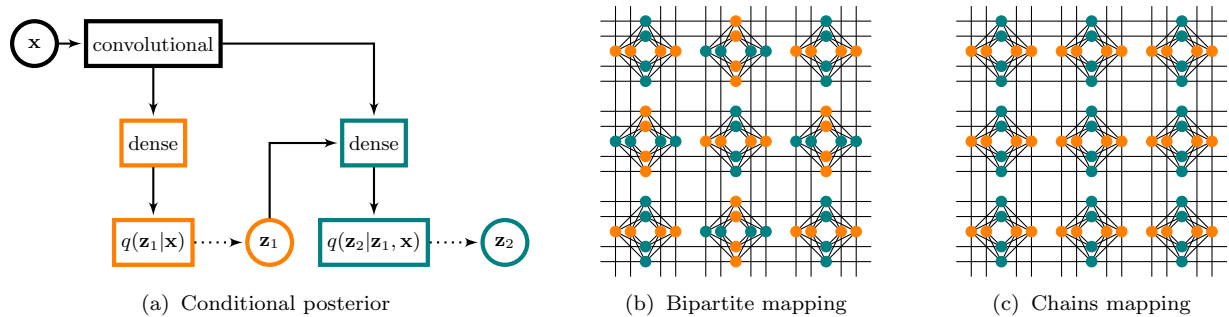

In our implementation, we have used fairly conventional convolutional neural networks. As we will discuss in this section, we have implemented only one specific architecture-dependent element in the encoder network that turned out to be very effective at improving performance on the Chimera graph. In general, we believe more elaborate, hardware-specific implementations of both the approximating posterior $q_{\bphi}(\z|\x)$ and the marginal distribution $p_{\btheta}(\x|\z)$ will significantly improve performance of VAE trained with analog devices with quasi two-dimensional connectivities. This is not just a problem of building a model with more capacity. As we discussed in the case of latent-unit use, it is also a problem of optimizing the model hyperparameters. When working with sparsely connected RBMs, it is easier for the KL term to push both approximating posterior and RBM prior to a local minimum of the loss function in which they are both trivial. Developing hardware-specific hybrid models will thus also aim at reaching local minima during training in which the RBM prior is as expressive as possible, exploiting the largest amount of correlations between latent units. 

In the context of hybrid VAE models trained with quantum annealers, the considerations above could replace more standard mapping techniques used in the quantum annealing community such as minor embedding~\cite{choi2008minor,choi2011minor} and majority voting. The latter techniques typically require a hard-coded specification of the hardware connectivity of each processor, which makes adapting the code to different processors cumbersome. Our perspective in the contest of hybrid generative modeling is to work by adapting classical networks using a high-level specification of the connectivity and letting the model, through stochastic gradient descent, learn the details of the connectivity of each processor (such as the locations of uncalibrated qubits).

We now discuss an example of how the classical networks can be optimized for a given architecture (in our case the Chimera graph). A common technique to implement a more expressive approximating posterior $q_{\bphi}(\z|\x)$ (with a tighter variational bound) is to introduce conditional relationships among latent units, also called hierarchies. In the case of two hierarchies, we first define an approximating posterior for a subset of latent units $\z_1$ and sample from it, then define a second approximating posterior for the remaining latent units $\z_2$ that depends conditionally on both the input data and the sampled values of the first group of latent variables:
\bea
\z_1 \sim q_{1,\bphi}(\z_1|\x)\,, \quad \z_2 \sim q_{2,\bphi}(\z_2,|\bzeta_1,\x)\,.
\label{eq:hier}
\eea
The schematic of our implementation is shown in Fig.~\ref{fig:7a}. We notice that models with a large number of hierarchies (possibly as large as the number of latent units) are possible. Such models, also referred to as autoregressive models~\cite{oord2016pixel}, are very powerful but have inefficient inference, since sampling must be performed sequentially and cannot be parallelized on modern GPU hardware.

The two hierarchical groups $(\z_1, \z_2)$, as well as the physical qubits on the Chimera connectivity, are not equivalent. We can thus build different models by simply choosing the mapping between the two hierarchical groups and an arbitrary bipartition of the physical qubits. Notice that training and deploying each of these models will involve exactly the same amount of classical and quantum computational resources. In our experiments we consider two possible mappings of the two hierarchical groups onto the physical qubits of the Chimera graph, which we call ``Bipartite" and ``Chains". The Bipartite mapping corresponds to the bipartite structure of the Chimera graph (see Fig.~\ref{fig:7b}). The Chains mapping corresponds to the vertical and horizontal physical layout of qubits of the Chimera architecture (see Fig.~\ref{fig:7c}). The identification of vertical and horizontal chains of qubits is commonly used to perform a minor embedding of a fully connected RBM on the Chimera graph. We stress again, however, that we never perform any minor embedding, and we always sample from the native Chimera graph in all cases.

\begin{figure*}[t]
\begin{center}
\includegraphics[width=1\columnwidth]{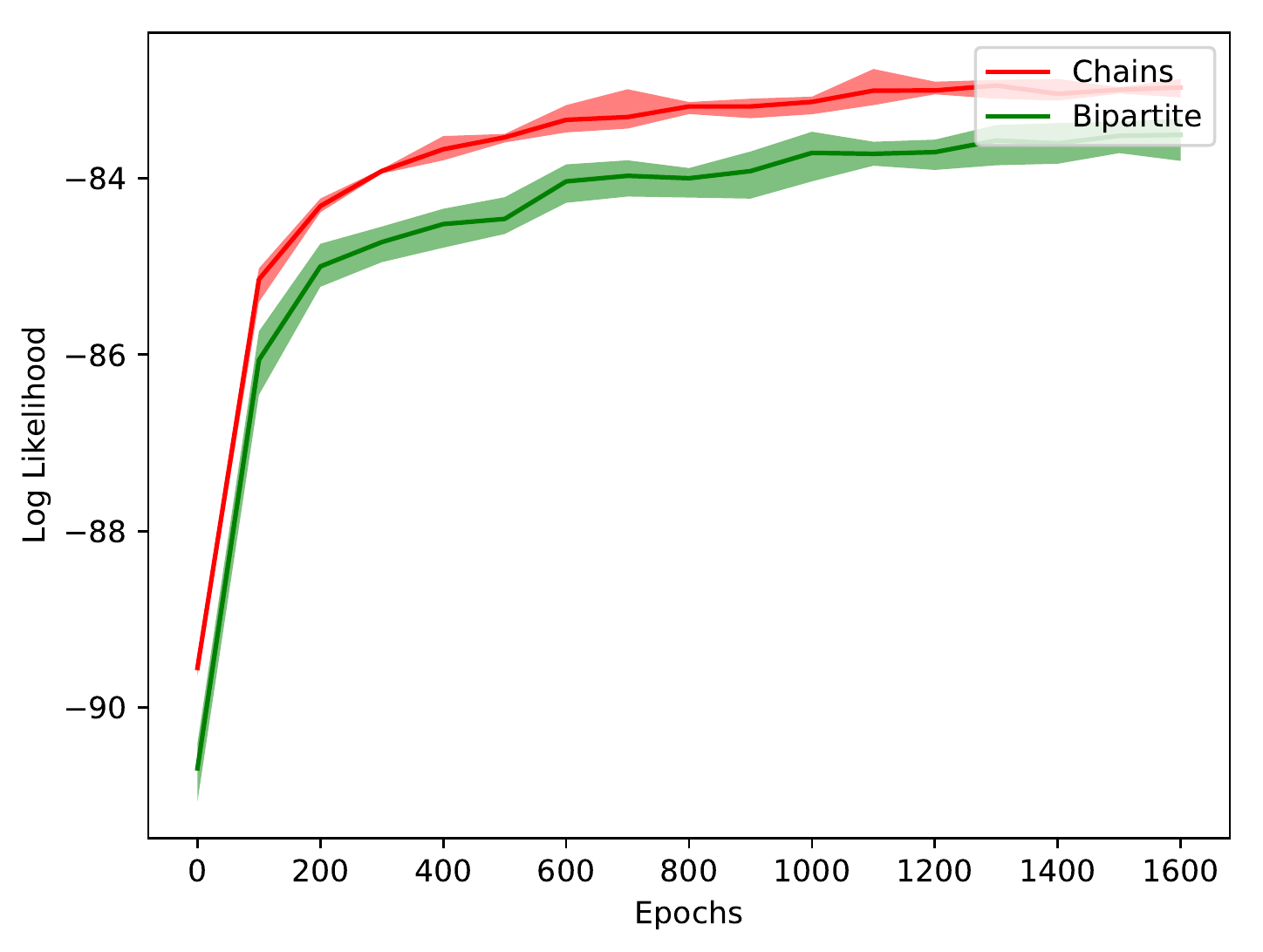}
\includegraphics[width=1\columnwidth]{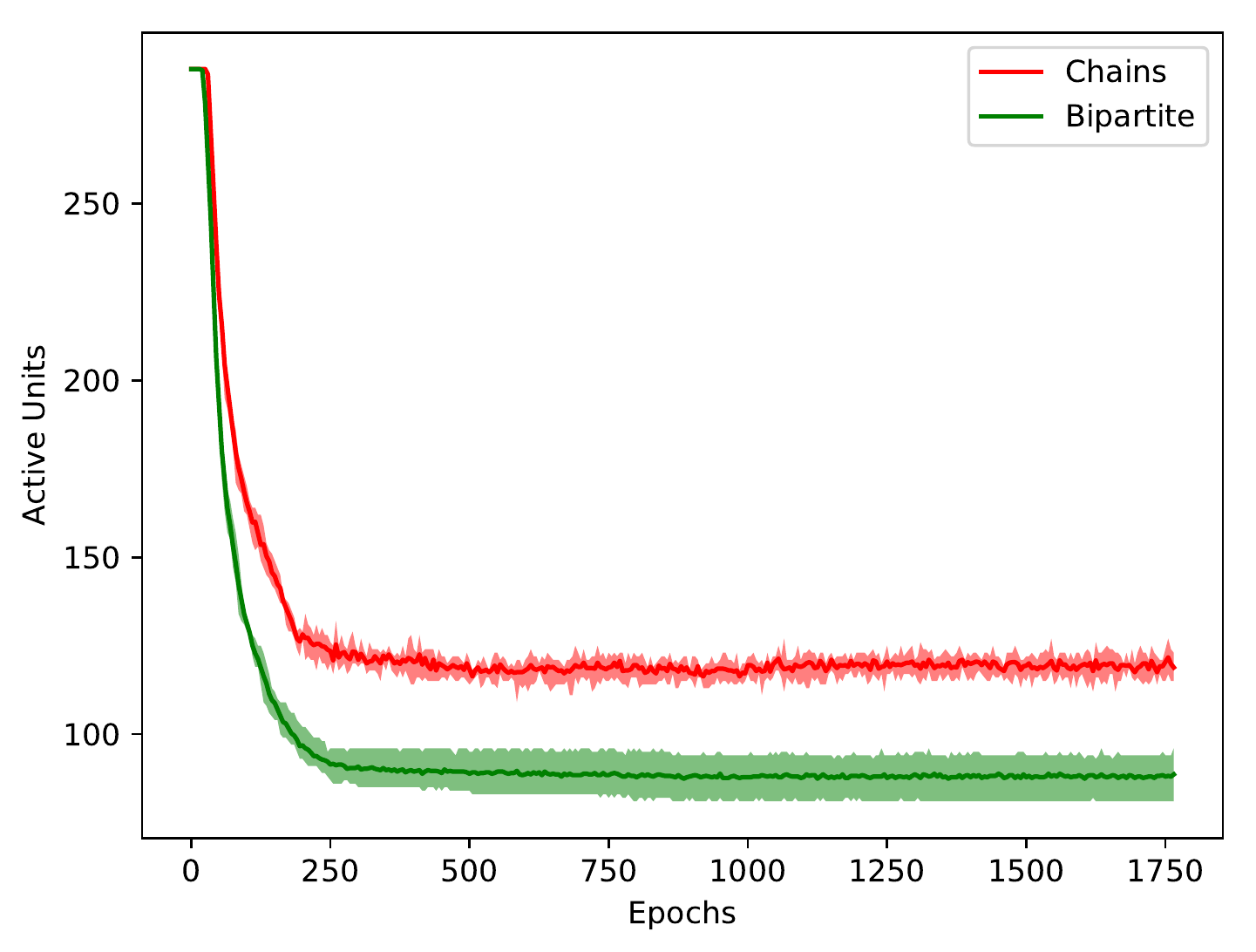}
\caption{Left panel: Log-likelihood comparison for chains and bipartite mappings. Right panel: Active units comparison for chains and bipartite mappings.}
\label{fig:8}
\end{center}
\end{figure*}

Comparative results of the two mappings, obtained with classical sampling, are shown in Fig.~\ref{fig:8}. In the left panel we see that there is a sizeable difference in generative performance between the two mappings, with the Chains mapping performing remarkably better. This better performance is also reflected in the much higher number of latent units exploited by the Chains mapping (see right panel of  Fig.~\ref{fig:8}). An intuitive explanation of the results above can be given as follows. Let us first write the KL term as:
\bea
D_{KL}( q_{\bphi}(\z|\x) ||  p_{\btheta}(\z)) & = &  D_{KL}( q_{1,\bphi}(\z_1|\x) ||  p_{\btheta}(\z_1)) + \nn
&+& D_{KL}( q_{2,\bphi}(\z_2|\z_1, \x) ||  p_{\btheta}(\z_2|\z_1))\,.\nn
\label{eq:KL-maps}
\eea 
In the Bipartite mapping, the conditional $p_{\btheta}(\z_2|\z_1)$ has a simple form that can be computed analytically due to the bipartite structure of the Chimera RBM. During training, it is very easy for the model to use the capacity of $q_{2,\bphi}(\z_2|\z_1, \x)$ to match the simple prior marginal $p_{\btheta}(\z_2|\z_1)$ and be independent from $\x$. As a consequence, a large portion of the representational capacity of the approximating posterior is wasted in representing the simple marginal $p_{\btheta}(\z_2|\z_1)$. In the Chains mapping, in contrast, the marginal $p_{\btheta}(\z_2|\z_1)$ is nontrivial, and the approximating posterior $q_{2,\bphi}(\z_2|\z_1, \x)$ has more difficulties in matching it and decoupling $\x$. As a consequence, the model ends up using more efficiently all the variational parameters $\bphi$. This is another manifestation of the optimization challenge present with VAE models mentioned before, which in this case it is exploited to find better local minima of the loss function.

In this section we have shown how a simple architecture-aware modification of the encoder network allows us to train better models and to exploit a given architecture more efficiently. We expect that architecture-aware model-engineering will be crucial to fully exploit large physical connectivities in the latent space of VAE.

%%%%%%%%%%%%%%%%%%%%%%%%%%%%%%%%%%%%%
\subsubsection{Training on larger datasets}
\label{sec:FMNIST}
%%%%%%%%%%%%%%%%%%%%%%%%%%%%%%%%%%%%%

Implementing more highly connected RBMs and developing classical encoders and decoders tailored to a given connectivity can only go so far in helping to exploit larger latent spaces. Together with other techniques such as KL-term anneal, the ideas mentioned in the previous two sections help reduce the pressure of the KL term to reach suboptimal local minima. In essence, VAE are also efficient lossy encoders. An alternative direction to increase latent space utilization is thus to train on more complex datasets. By doing so, a larger number of latent units is necessary to store enough information such that the reconstruction term (first term in Eq.~\ref{eq:ELBO_split}) is large enough.

\begin{figure*}[t]
\begin{center}
\includegraphics[width=1\columnwidth]{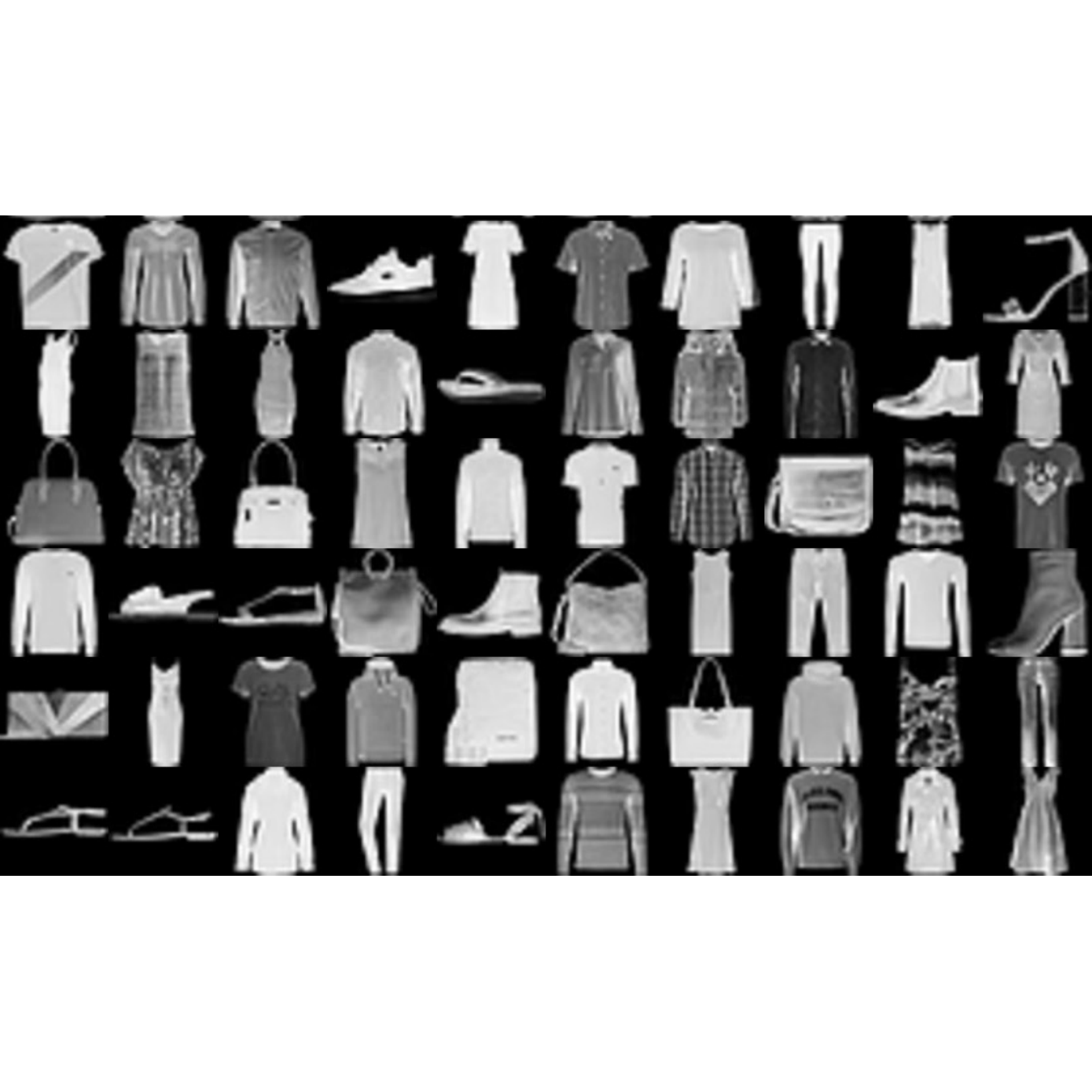}
\includegraphics[width=1\columnwidth]{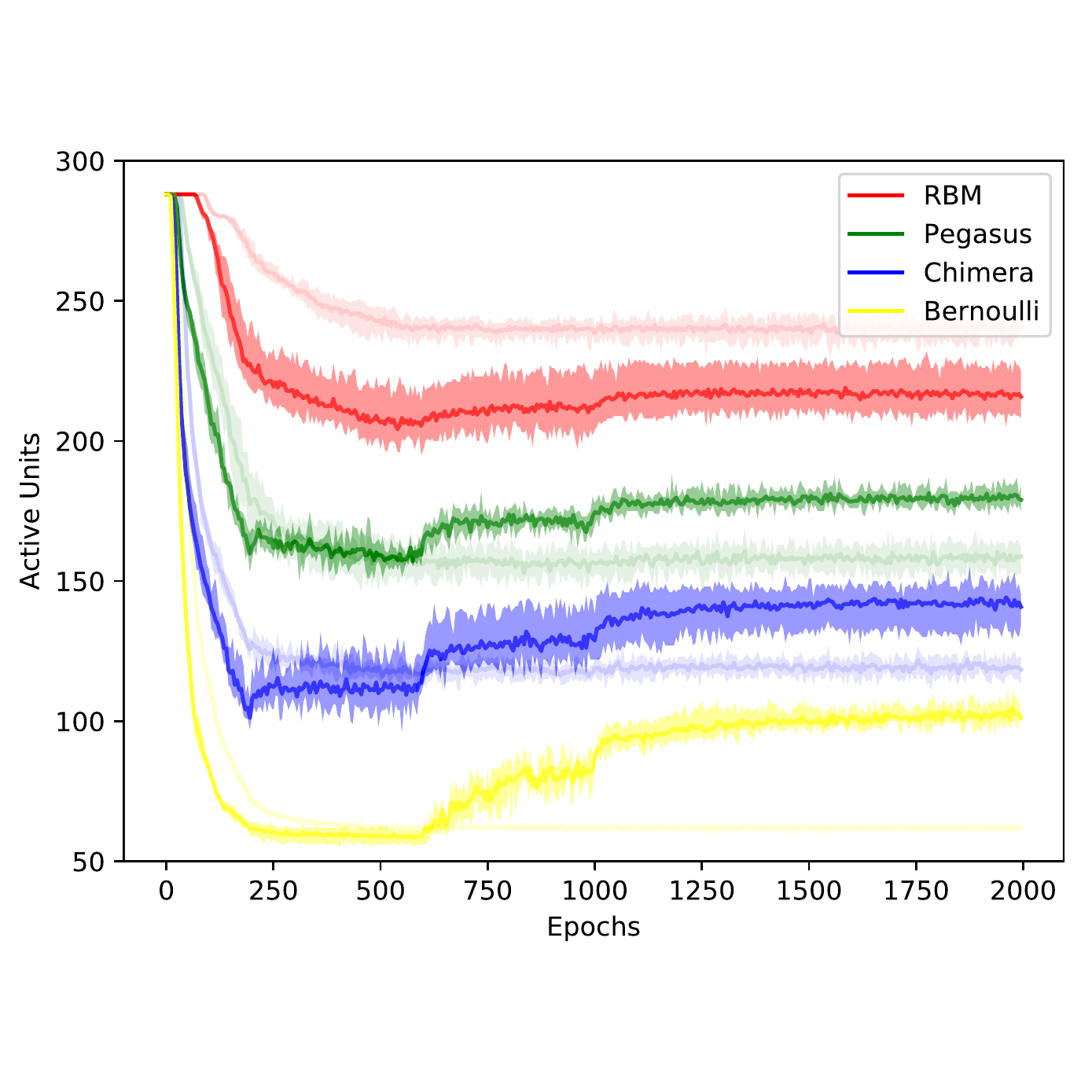}
\caption{Left panel: Fashion MNIST dataset. Right panel: Active units for the same VAE models trained on FMNIST (solid colors) and on MNIST (shaded colors). All models with sparse latent connectivities use a much larger number of latent units when trained on the more complex dataset.}
\label{fig:9}
\end{center}
\end{figure*}

We give numerical evidence of the intuition above by training the same VAE models used in the previous sections on the Fashion MNIST (FMNIST) dataset. A set of images from the FMINST dataset is shown in the left panel of Fig.~\ref{fig:9}. FMNIST is the same size as MNIST (50000, 28$\times$28 images) and has the same number of classes. However, its images are more complex with more fine details, including grey-scale features that are important for correct image classification (whereas MNIST digits are substantially black and white). In the right panel of Fig.~\ref{fig:9}, we train the same models on MNIST (shaded) and FMNIST (solid) and compare the number of active units during training. We see that, apart from the case with fully connected RBMs, all other models use a substantially larger number of latent units.

%%%%%%%%%%%%%%%%%%%%%%%%%%%%%%%%%%%%%
\subsection{Multi-modality of latent-space RBMs}
\label{sec:MULTI}
%%%%%%%%%%%%%%%%%%%%%%%%%%%%%%%%%%%%%

Exploiting large latent-space RBMs is a necessary condition to eventually achieve quantum advantage when sampling with quantum annealers. This condition is however not sufficient. The typical computational bottleneck in training an RBM is due to the appearance of well-defined modes. These modes make classical sampling techniques inefficient and slow-mixing, resulting in highly correlated samples used both during training and generation. While making sampling harder, the development of multi-modal distributions is actually an appealing property of RBMs, since it allows such models to represent complex and powerful probability distributions. The idea behind searching for quantum advantage in training RBM with quantum annealers is, indeed, to exploit quantum resources (such as tunneling) to more efficiently mix between different modes in the landscape defined by the RBM.

When trained on visible data, an RBM naturally develops complex landscapes to match the complexity of the statistical relationship present in the training data. However, while RBMs trained on latent representations can potentially develop well-defined modes, they do not necessarily do so. In fact, one of the capabilities of generative models with latent variables is finding a set of statistically independent latent features~\cite{dinh2014nice}. This is typically enforced during model building by using trivial priors such as the product of independent Gaussian (for continuous latent units) or the product of independent Bernoulli (for discrete latent spaces). Even when the prior is potentially complex and trainable, as an RBM, the presence of the KL term can push the model during training into local minima in which the trained RBM develops a trivial landscape.

\begin{figure*}[t]
\begin{center}
\subfigure[\, Bernoulli.]{\includegraphics[width=0.5\columnwidth]{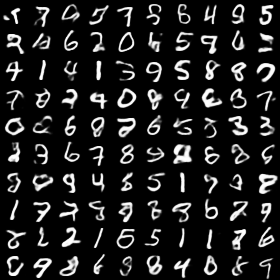}\label{fig:10a}}
\subfigure[\, Chimera.]{\includegraphics[width=0.5\columnwidth]{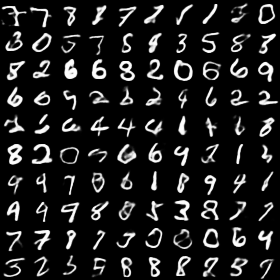}
\label{fig:10b}}
\subfigure[\, Pegasus.]{\includegraphics[width=0.5\columnwidth]{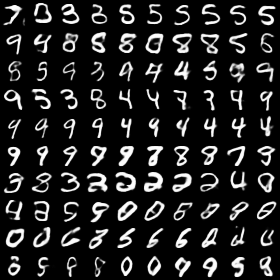}
\label{fig:10c}}
\subfigure[\, RBM.]{\includegraphics[width=0.5\columnwidth]{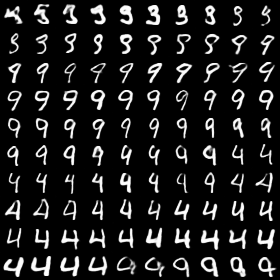}
\label{fig:10d}}\\
\caption{Block Gibbs sampling with different connectivities. Going from left to right, denser connectivities result in more well-defined modes developed in the trained RBM. Especially in the case of Pegasus and an RBM, for example, it is clearly visible how the block Gibbs chain is trapped in a typical basin of the landscape for MNIST connected to the digits 4 and 9.}
\label{fig:10}
\end{center}
\end{figure*}

In this section we give evidence that RBMs trained in the latent space of a VAE model do indeed develop a nontrivial landscape with well-defined modes. As we have shown in Sec.~\ref{sec:CONN} (see Fig.~\ref{fig:6}), RBMs with denser connectivities naturally lead to better performing VAEs. This is an indirect indication that we are indeed exploiting the additional capacity and expressivity of more connected RBMs. In this section we give more explicit evidence of this. In Fig.~\ref{fig:10}, we generate a sequence of images via block Gibbs sampling. The top left image is generated by picking a latent configuration $\z$ out of a uniform distribution over all configurations. This latent sample is then sent through the decoder. Going from left-to-right, top-to-bottom, each subsequent image is obtained after updating all latent units with a sequence of block Gibbs updates (one for Bernoulli, two for the bipartite connectivities Chimera and RBM, four for the quadripartite Pegasus connectivity). As expected, in the Bernoulli case (Fig.~\ref{fig:10a}), each update results in uncorrelated samples. The Chimera connectivity (Fig.~\ref{fig:10b}) is able to develop weakly correlated samples, as shown by short sequences of similar images. Correlated samples with well-defined modes are more clearly visible with the Pegasus connectivity ((Fig.~\ref{fig:10c})). Finally, we confirm that increasing the connectivity up to a fully connected RBM (Fig.~\ref{fig:10d}) results in long sequences of correlated samples and related to the deep valleys of the RBM energy landscape.

The results shown in Fig.~\ref{fig:10} show that RBMs trained as priors of generative models with latent variables naturally learn multi-modal, nontrivial probability distributions. These distributions are expressive, making the whole VAE more expressive, while at the same time developing the same types of computational bottlenecks that make classical sampling algorithms inefficient. This paves the way to effectively use quantum annealers as means to more scalable quantum-assisted sampling, enabling us to sample from RBMs of sizes and complexity that are infeasible with classical methods.

\begin{figure*}[t]
\begin{center}
\subfigure[\, Models trained on different D-Wave 2000Q achieved performance comparable to training with population annealing. Log-likelihood evaluation for Baseline D-Wave 2000Q did not converge due to diverging weights.]{\includegraphics[width=1\columnwidth]{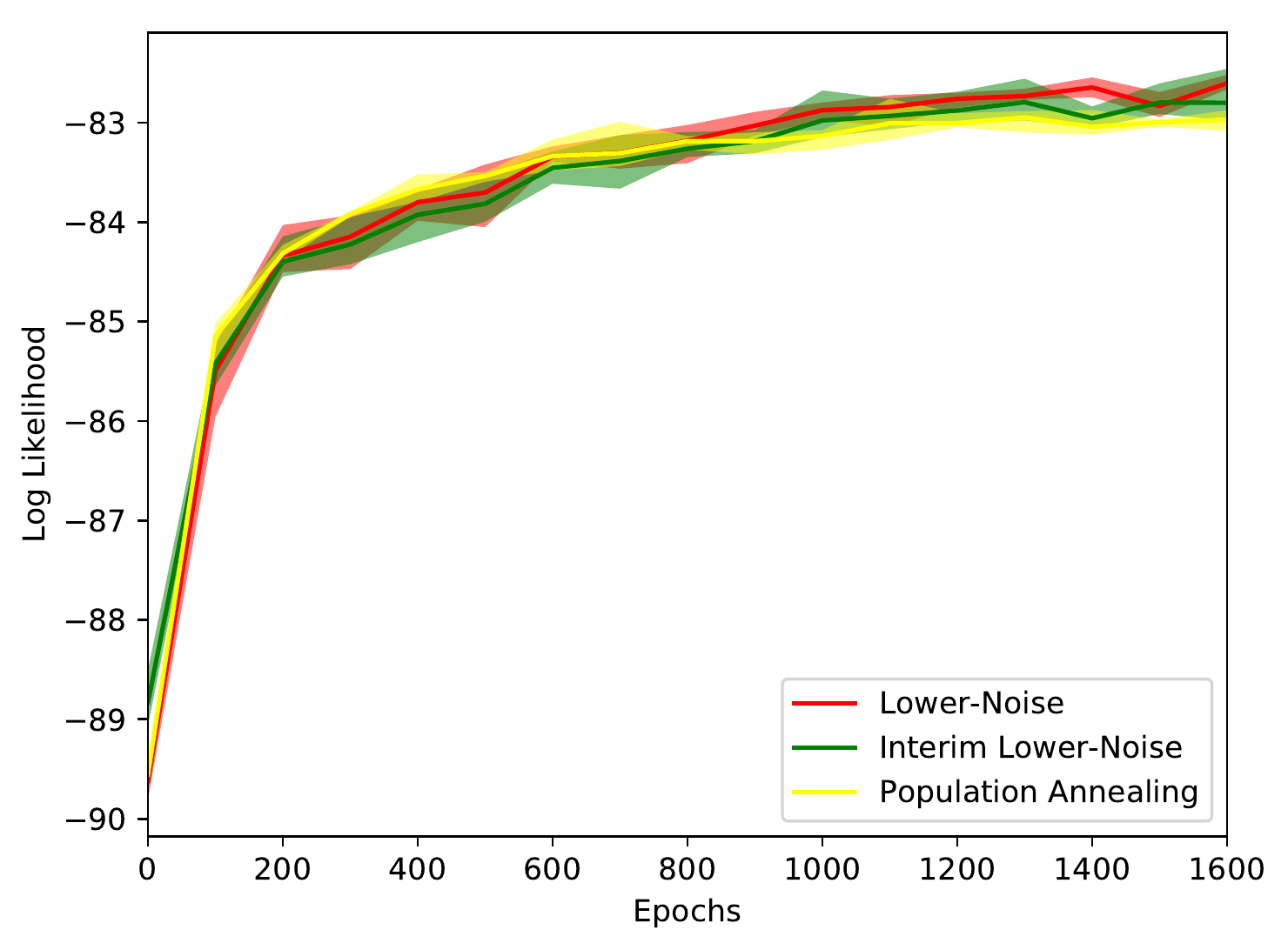}\label{fig:11a}}
\subfigure[\, The weights of the Baseline D-Wave 2000Q weights diverge after about 100 epochs. For the Interim Lower Noise processors weights are suboptimally small. Weight size of the Lower Noise D-Wave 2000Q is closest to population annealing.]{\includegraphics[width=1\columnwidth]{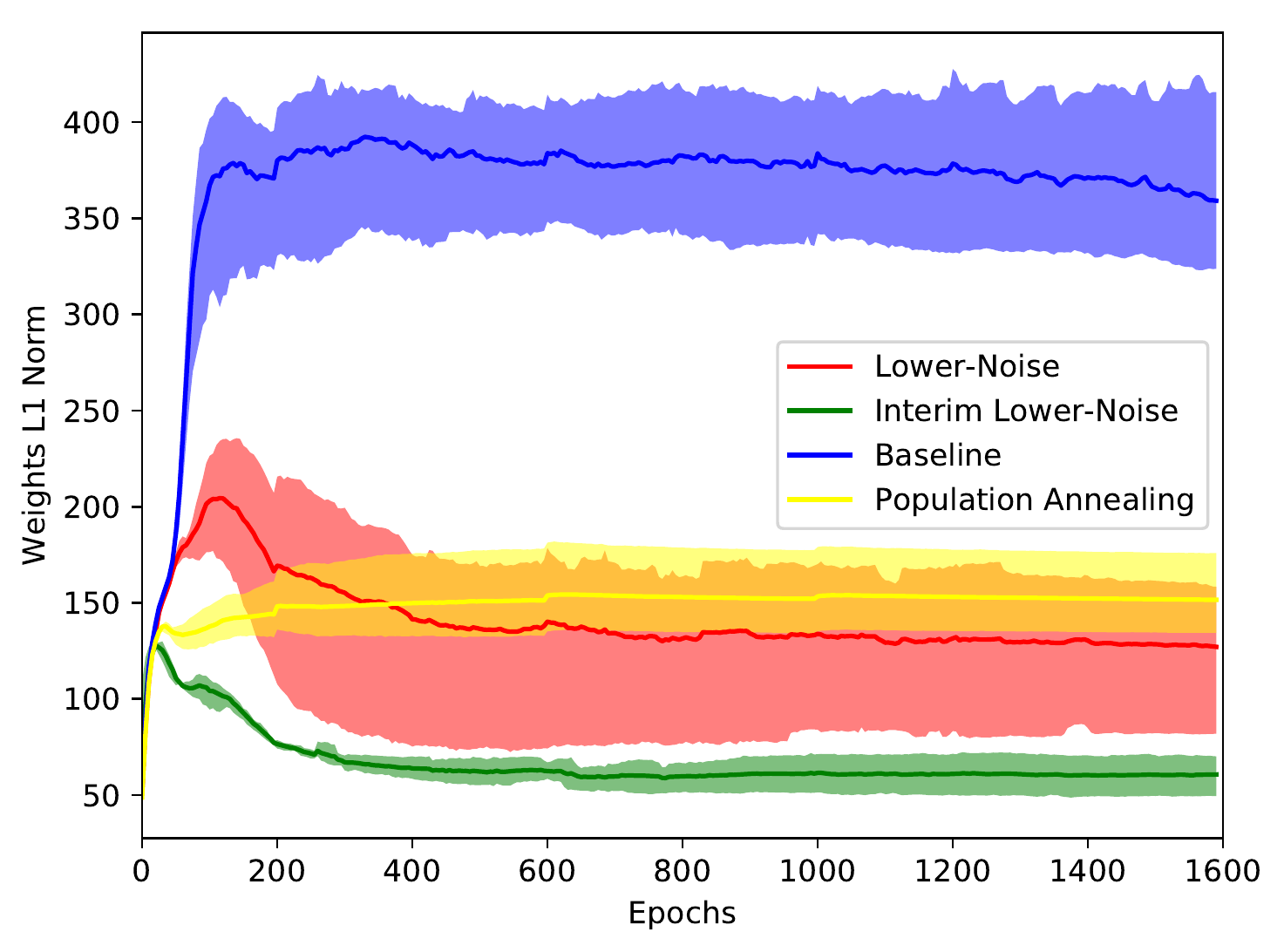}\label{fig:11b}}\\
\subfigure[\, The Lower-Noise D-Wave 2000Q is able to exploit a larger number of latent units than the more noisy Interim Lower-Noise D-Wave 2000Q.]{\includegraphics[width=1\columnwidth]{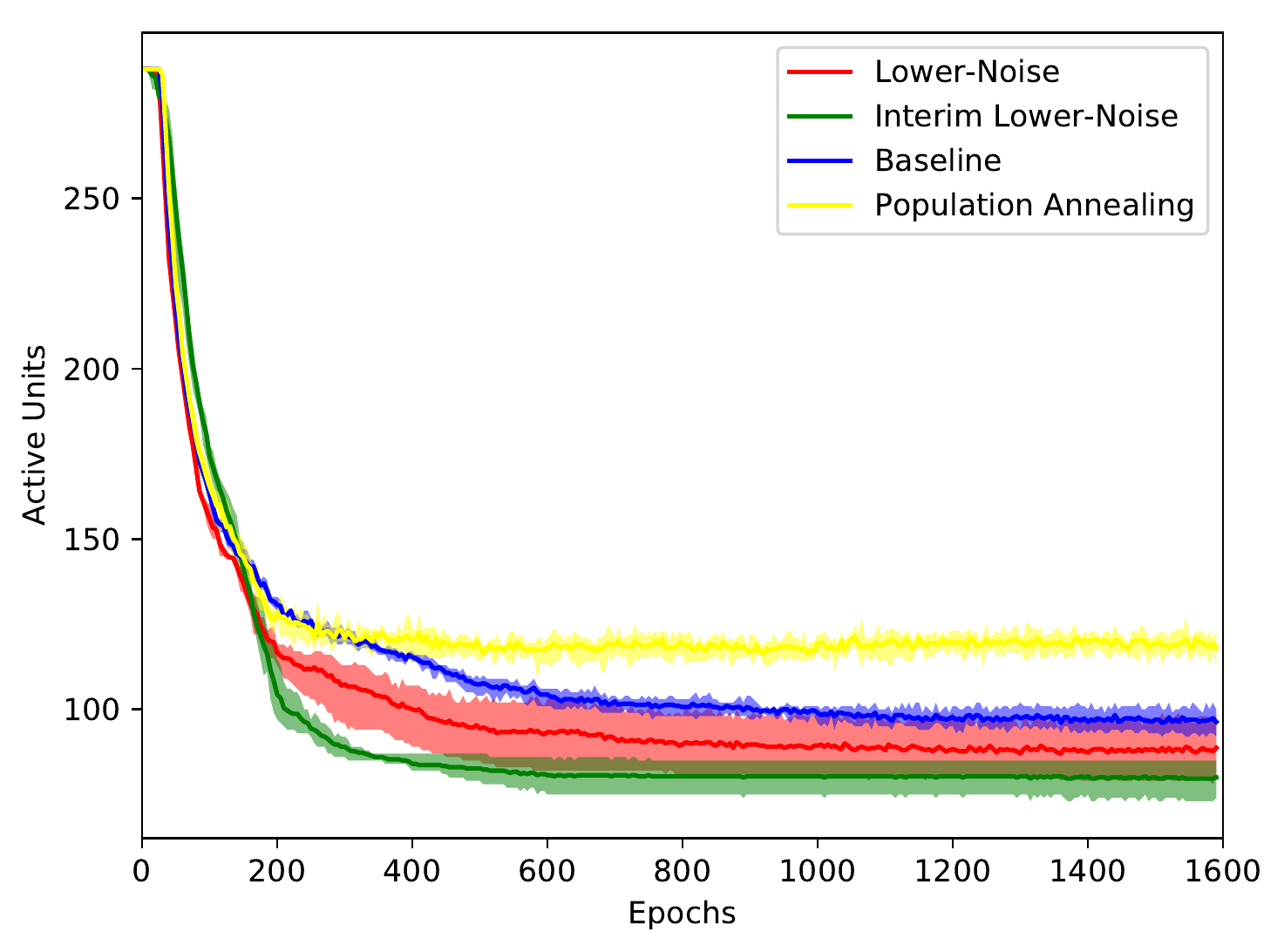}\label{fig:11c}}
\subfigure[\, Different temperature profiles during training  for the three D-Wave 2000Q.  ]{\includegraphics[width=1\columnwidth]{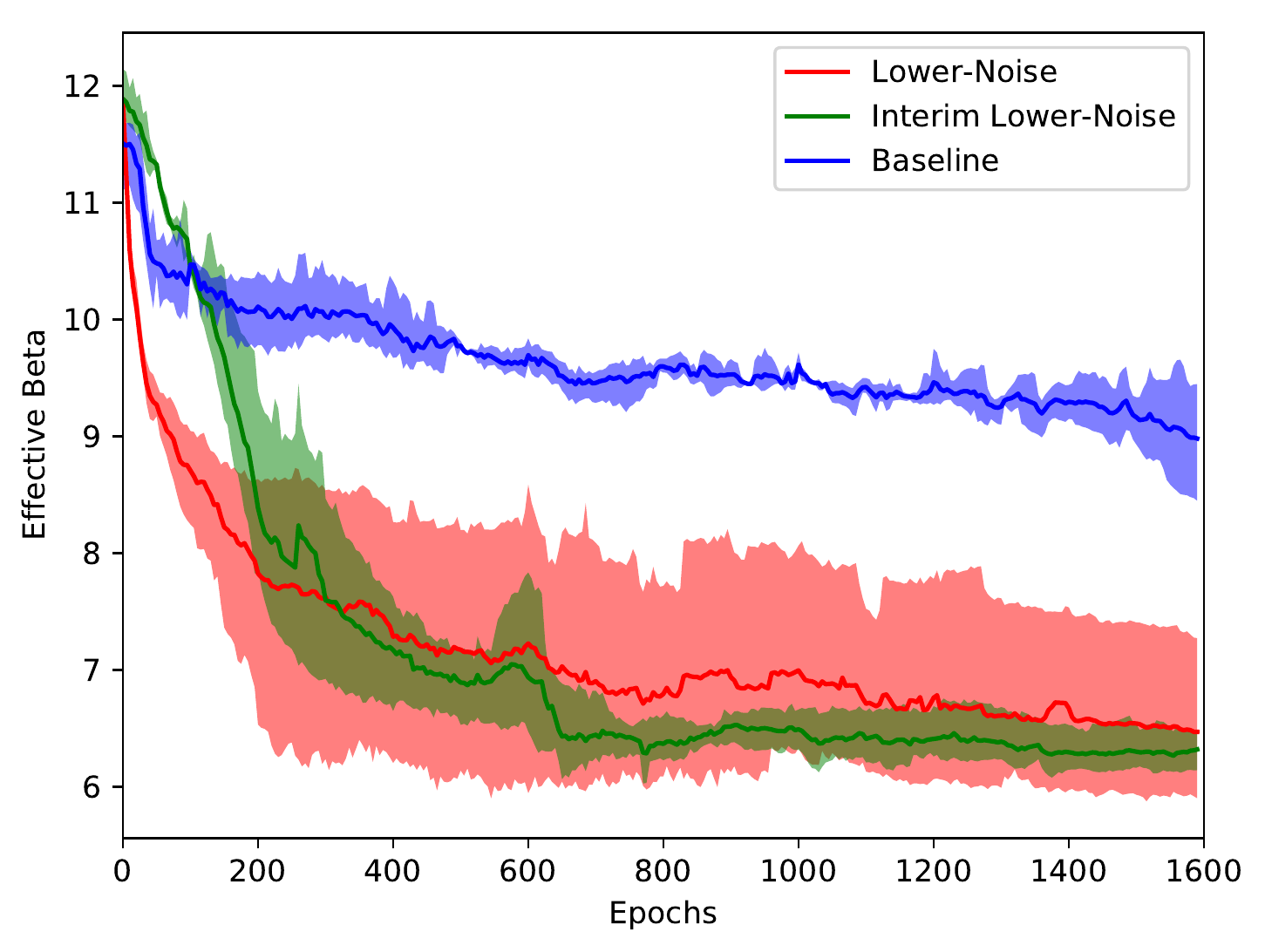}\label{fig:11d}}
\caption{Training on different quantum annealers with different noise profiles.}
\label{fig:11}
\end{center}
\end{figure*}

%%%%%%%%%%%%%%%%%%%%%%%%%%%%%%%%%%%%%
\subsection{Robustness to noise and control errors}
\label{sec:NOISE}
%%%%%%%%%%%%%%%%%%%%%%%%%%%%%%%%%%%%%

Using quantum annealers to train large RBMs directly on complex data remains challenging. Apart from the unsatisfying performance of using RBMs with quasi two-dimensional connectivities on visible data, a major difficulty is biased sampling (and thus inaccurate gradients) obtained with quantum annealers. There are two main sources of bias: control errors and imperfect or incomplete thermalization at the freezing point. While the latter can be improved with appropriate pause-and-ramp annealing schedules, the former can only be improved with technological advancements. Despite these known difficulties, we have shown in the previous sections we have successfully trained large RBM (hundreds of units) in the latent space of a VAE solely using samples coming from a D-Wave 2000Q, without using any hard-coded pre or post-processing to the raw data obtained from the annealer.

We interpret our positive results as an indication that training RBMs with our setup is relatively robust to noise and control errors. In fact, we can interpret both the encoder and decoder as powerful tools to pre- and post-process data to be sent to the quantum annealer. Using stochastic gradient descent, we train the encoder and decoder to generate a set of latent features that are more easily modeled by the latent-space RBM. For example, real images might have strongly correlated, sharp features (such as regions with black or white pixels), which require large weights to be modeled correctly. A precise implementation of such large weights might be challenging for analog devices with finite range such as quantum annealers. Additionally, both encoders and decoders might be able to learn and correct, or at least reduce the effects of, systematically biased sampling.

To investigate the role of noise and control errors in determining sampling quality and performance of the trained models, we perform a set of comparative experiments in which we train the same model on MNIST dataset, using samples coming from three D-Wave 2000Q with different noise profiles. The Baseline and Lower-Noise D-Wave 2000Q are both publicly available on D-Wave's Leap\textsuperscript{TM} cloud service. We have also included an Interim Lower-Noise processor with an intermediate noise profile that is internally available at D-Wave. 

Results are shown in Fig.~\ref{fig:11}. In Fig.~\ref{fig:11a} we report the log-likelihood during training. Models trained on the Lower and Interim Lower-Noise D-Wave 2000Q achieved performance comparable to training with population annealing. The evaluation of the log-likelihood for the Baseline D-Wave 2000Q diverged due to diverging weights, as can be seen in Fig.~\ref{fig:11b}. The weights of the Baseline processor start diverging after about 100 epochs.  The Interim Lower-Noise processor shows an opposite behavior, with weights getting small and plateauing after about 500 epochs.  For the Lower-Noise processor, the L1 of the weights plateaus at a value that is closer to that obtained with ``noiseless" population annealing. Notice that a consistent comparison in Fig.~\ref{fig:11b} we have reported the weights $W$ rescaled by the effective temperature (see Eq.~\ref{eq:freez}, and not the ``bare" annealing values $J$.  Figure~\ref{fig:11b} highlights the remarkably different response of three different quantum annealers to our model. Despite such differences, the performance of our hybrid implementation is robust (as shown in Fig.~\ref{fig:11a}) and does not require any hardware-specific adaptations or fine-tuning. Only while training with the Baseline D-Wave 2000Q, we needed a more aggressive clipping (that is restricting the weights and biases to have narrower range than the maximum allowed) to achieve similar performance and converged log-likelihood estimation.

Noise and control errors also manifest in a less efficient use of the latent space, as seen in Fig.~\ref{fig:11c}. All models trained with D-Wave 2000Q use fewer active units than population annealing. Since the estimate of the log-likelihood of the models trained with the Baseline processor did not converge, we focus on the comparison between the Interim and Lower Noise processor: the latter can exploit a larger number of latent units. We finally show in Fig.~\ref{fig:11d} the profile for the extracted effective temperature during training, which is remarkably different for the three D-Wave 2000Q. The results shown in Fig.~\ref{fig:11d} underlines the importance of developing advanced annealing techniques to stabilize temperature during training.

In this section we have demonstrated the robustness to noise of our implementation, as highlighted in Fig.~\ref{fig:11a}. At the same time, Fig.~\ref{fig:11c} shows an important effect of noise, which is to make it harder for our hybrid model to exploit the optimal number of latent units. We thus anticipate that exploiting large RBMs (with thousands of units, eventually) following the directions indicated in the previous sections must be accompanied  by continued efforts in reducing sampling bias due to noise and control errors of future-generation quantum annealing devices.

%%%%%%%%%%%%%%%%%%%%%%%%%%%%%%%%%%%%%
\section{Conclusions}
%%%%%%%%%%%%%%%%%%%%%%%%%%%%%%%%%%%%%
\label{sec:conc}
In this work, we have demonstrated the use of quantum annealers as  Boltzmann samplers to estimate the negative phase of classical RBMs placed in the latent space of deep convolutional variational autoencoders. This setup allows for the construction of quantum-classical hybrid generative models that can be scaled to large, realistic datasets. We have mostly experimented with MNIST, a common testbed dataset which includes $50,000$, $28\times 28$ binarized handwritten digits to achieve a log-likelihood of about $-82.2\pm0.2$ nats, which compares favorably to state-of-the-art achieved with autoregressive models ($-78.5$ nats (natural unit of information)). In addition to demonstrating scalability and performance, we have discussed several other features of our hybrid approach.

First, we are able to use quantum annealers as ``native samplers", that is, samplers from their native graph: we do not use any hard-coded encoding-decoding scheme such as minor embedding or majority vote. Arguably this is one of the most effective ways to exploit the computational capabilities of quantum annealers. The encoding and decoding process is indeed efficiently performed by deep convolutional networks, which are trained to extract relevant feature via stochastic gradient descent. As we have shown, this approach is particularly flexible, since it naturally adapts to different connectivities and arbitrary working graphs. 

Second, by successfully training the same model on three quantum annealers with  different noise profiles, we have shown that our implementation is fairly robust to noise and control errors. Indeed, the deep convolutional networks can be seen as learned pre- and post-processing steps that regularize both the visible data and the effects of noise. A key reason of the success of our implementation is indeed the fact that the weights and biases as implemented on the quantum annealers rarely grow (during training) beyond their allowed range, even with minimal or no regularization. This result is to be contrasted to training RBMs directly on visible data, for which weights are typically much larger and regularization is critical to avoid overfitting. The latter case is much more challenging for analog devices with limited allowed range.

The quantum-classical hybrid models we have considered in this work employ a large amount of classical computing power performed on modern GPUs. The computational task that we offloaded to the quantum annealer (sampling from the latent-space RBMs) can still be performed classically at a fraction of the overall computational cost. To achieve any form of quantum advantage in this framework, we need to offload generative capacity to the prior, by exploiting large RBMs capable of representing complex probability distributions from which classical sampling becomes too expensive. We have provided evidence that this path to quantum advantage is possible by deploying annealers with denser connectivities and lower noise, engineering classical neural nets that better exploit physical connectivities and by working with more complex datasets. All these improvements seem achievable in the near future, and represent possible interesting lines of research that we leave for future work.

%%%%%%%%%%%%%%%%%%%%%%%%%%%%%%%%%%%%%
\subsection*{Acknowledgements}
%%%%%%%%%%%%%%%%%%%%%%%%%%%%%%%%%%%%%
The authors would like to thank Arash Vahdat and Jason T. Rolfe for useful discussions during the preparation of this work, Kelly Boothby for providing the figures for Pegasus and Chimera architectures, and Fiona Hanington for editing the manuscript. For this work L.B. was supported by a grant from ``Fondazione Angelo Della Riccia". W.V is grateful for support from NASA Ames Research Center, the Office of the Director of National Intelligence (ODNI) and the Intelligence Advanced Research Projects Activity (IARPA), via IAA 145483. We also appreciate support from the AFRL Information Directorate under grant F4HBKC4162G001. The views and conclusions contained herein are those of the authors and should not be interpreted as necessarily representing the official policies or endorsements, either expressed or implied, of IARPA, AFRL, or the U.S. Government. The U.S. Government is authorized to reproduce and distribute reprints for Governmental purpose notwithstanding any copyright annotation thereon.

\bibliography{refs}
%%%%%%%%%%%%%%%%%%%%%%%%%%%%%%%%%%%%%

\appendix
\label{sec:APP}
%%%%%%%%%%%%%%%%%%%%%%%%%%%%%%%%%%%%%

\section{Convolutional VAE}
\label{sec:CONVVAE}
%%%%%%%%%%%%%%%%%%%%%%%%%%%%%%%%%%%%%
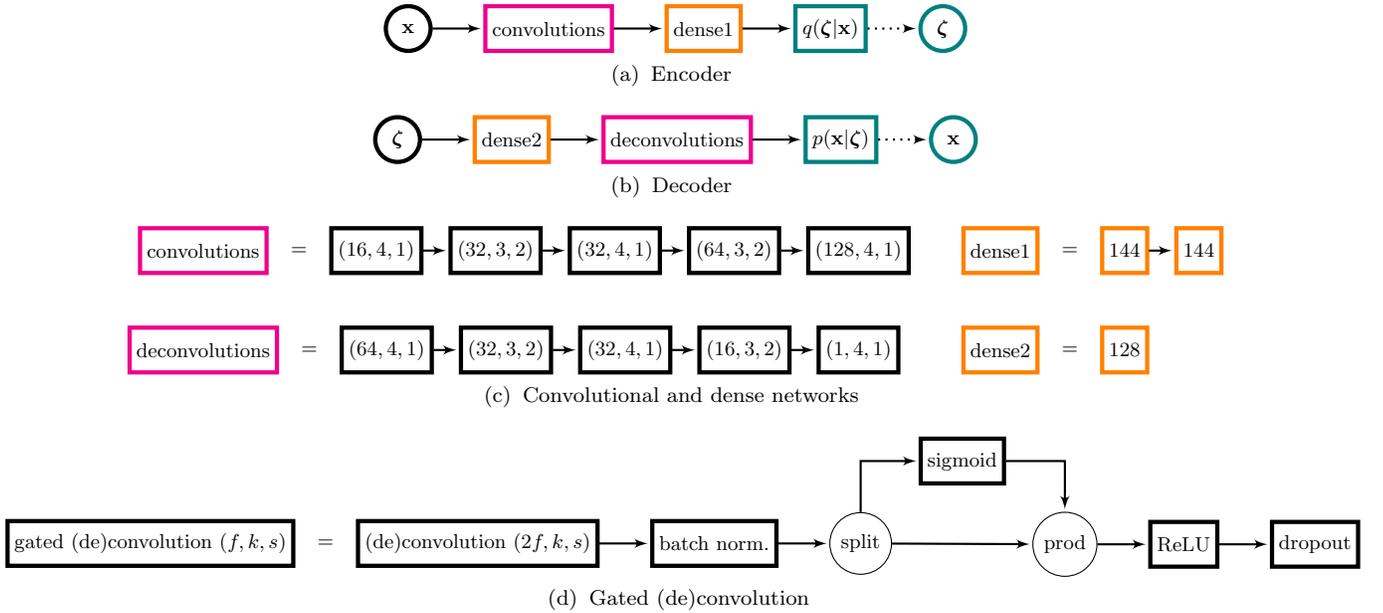
\begin{figure*}[t]
 \centering
 \subfigure[\, Encoder]{
  \begin{tikzpicture}[->,>=stealth',shorten >=1pt,auto,node distance=1.0cm, thick, scale=0.85]
       \tikzstyle{every state}=[fill=white,draw=black,text=black, transform shape, shape=rectangle, minimum size=0.7cm, ultra thick]
       \tikzstyle{line} = [draw, -latex']
       \node at (0,0) [state, shape=circle]                      (x)           {$\x$};
       \node [state, draw=magenta, right=0.8 of x]                  (conv)        {convolutions};
       \path [line] (x) -- (conv);
       \node [state, draw=orange, right=0.8 cm of conv]            (d2)         {dense1};
       \node [state, draw=teal, right=0.8 cm of d2]            (r2)         {$q(\bzeta|\x)$};
       \node [state,  draw=teal, shape=circle, right=0.8 cm of r2]          (z2)         {$\bzeta$};
       \path [line] (d2) -- (r2);
       \path [line] (conv) --  (d2);
       \path [line, dotted] (r2) --  (z2);
  \end{tikzpicture}\label{fig:12a}}
  
   \subfigure[\, Decoder]{
  \begin{tikzpicture}[->,>=stealth',shorten >=1pt,auto,node distance=1.0cm, thick, scale=0.85]
       \tikzstyle{every state}=[fill=white,draw=black,text=black, transform shape, shape=rectangle, minimum size=0.7cm, ultra thick]
       \tikzstyle{line} = [draw, -latex']
       \node at (0,0) [state, shape=circle]          (z)           {$\bzeta$};
       \node [state, draw=orange, right=0.8 cm of z]            (d2)         {dense2};
       \node [state, draw=magenta, right=0.8 of d2]          (conv)        {deconvolutions};
       \path [line] (z) -- (d2);
       \node [state, draw=teal, right=0.8 cm of conv]            (r2)         {$p(\x|\bzeta)$};
       \node [state,  draw=teal, shape=circle, right=0.8 cm of r2]          (z2)         {$\x$};
       \path [line] (d2) --  (conv);
       \path [line] (conv) --  (r2);
       \path [line, dotted] (r2) --  (z2);
  \end{tikzpicture}\label{fig:12b}}
  
   \subfigure[\, Convolutional and dense networks]{
  \begin{tikzpicture}[->,>=stealth',shorten >=1pt,auto,node distance=1.0cm, thick, scale=0.85]
       \tikzstyle{every state}=[fill=white,draw=black,text=black, transform shape, shape=rectangle, minimum size=0.7cm, ultra thick]
       \tikzstyle{line} = [draw, -latex']
       \node at (0,0) [state, draw=magenta]          (conv)        {convolutions};
       \node [state,  draw=white, shape=circle, right=0.1 cm of conv]          (equal)         {$=$};
       \node [state, right=0.1 of equal]          (gate_1)        {$(16,4,1)$};
       \node [state, right=0.4 of gate_1]          (gate_2)        {$(32,3,2)$};
       \path [line] (gate_1) --  (gate_2);
       \node [state, right=0.4 of gate_2]          (gate_3)        {$(32,4,1)$};
       \path [line] (gate_2) --  (gate_3);
       \node [state, right=0.4 of gate_3]          (gate_4)        {$(64,3,2)$};
       \path [line] (gate_3) --  (gate_4);
       \node [state, right=0.4 of gate_4]          (gate_5)        {$(128,4,1)$};
       \path [line] (gate_4) --  (gate_5);
       
       \node [state, draw=orange, right=0.8 of gate_5]          (dense)        {dense1};
       \node [state,  draw=white, shape=circle, right=0.1 cm of dense]          (equal_0)         {$=$};
       \node [state, draw=orange, right=0.1 of equal_0]          (dense_11)        {$144$};
      \node [state, draw=orange, right=0.4 of dense_11]          (dense_12)        {$144$};
      \path [line] (dense_11) --  (dense_12);

       \node [state, draw=magenta, below=0.8 of conv]          (deconv)        {deconvolutions};
       \node [state,  draw=white, shape=circle, right=0.1 cm of deconv]          (equal_1)         {$=$};
       \node [state, right=0.1 of equal_1]          (gate_11)        {$(64,4,1)$};
       \node [state, right=0.4 of gate_11]          (gate_21)        {$(32,3,2)$};
       \path [line] (gate_11) --  (gate_21);
       \node [state, right=0.4 of gate_21]          (gate_31)        {$(32,4,1)$};
       \path [line] (gate_21) --  (gate_31);
       \node [state, right=0.4 of gate_31]          (gate_41)        {$(16,3,2)$};
       \path [line] (gate_31) --  (gate_41);
       \node [state, right=0.4 of gate_41]          (gate_51)        {$(1,4,1)$};
       \path [line] (gate_41) --  (gate_51);

       \node [state, draw=orange, below=0.8 of dense]          (dense3)        {dense2};
       \node [state,  draw=white, shape=circle, right=0.1 cm of dense3]          (equal_2)         {$=$};
       \node [state, draw=orange, right=0.1 of equal_2]          (dense_21)        {$128$};
  \end{tikzpicture}\label{fig:12c}}
  
  \subfigure[\, Gated (de)convolution]{
  \begin{tikzpicture}[->,>=stealth',shorten >=1pt,auto,node distance=1.0cm, thick, scale=0.85]
       \tikzstyle{every state}=[fill=white,draw=black,text=black, transform shape, shape=rectangle, minimum size=0.7cm, ultra thick]
       \tikzstyle{line} = [draw, -latex']
       \node at (0,0) [state]          (b)        {gated (de)convolution $(f, k, s)$};
       \node [state, shape=circle, draw=white, right=0.1 of b]          (bp)        {$=$};
       \node [state, right=0.1 of bp]          (b1)        {(de)convolution $(2 f, k, s)$};
       \node [state, right=0.8 of b1]          (b2)        {batch norm.};
       \path [line] (b1) --  (b2);
       \node [state, shape=circle, thin, right=0.8 of b2]          (b3)        {split};
       \path [line] (b2) --  (b3);
       \node [state, above right=0.8 of b3]          (b4)        {sigmoid};
       \path [line] (b3) |-  (b4);
       \node [state, shape=circle, thin, below right=0.8 of b4]          (b5)        {prod};
       \path [line] (b4) -|  (b5);
       \path [line] (b3) --  (b5);
       \node [state, right=0.8 of b5]          (b6)        {ReLU};
       \path [line] (b5) --  (b6);
       \node [state, right=0.8 of b6]          (b7)        {dropout};
       \path [line] (b6) --  (b7);
  \end{tikzpicture}\label{fig:12d}}
 \caption{Detailed specification of the networks employed in our experiments.}
 \label{fig:12}
\end{figure*}

The VAE employed in our experiments is schematically represented in Fig.~\ref{fig:12}. Both approximating posterior $q(\z|\x)$ (encoder) and marginal $p(\x|\z)$ (decoder) are  constructed using deep convolutional networks, see Figs.~\ref{fig:12a} and \ref{fig:12b}. Although not technically necessary, we use (approximately) mirror implementations for encoder and decoder. In the encoder, down-sampling is achieved by employing strided convolutions, while in the decoder up-sampling is similarly obtained  with strided deconvolutions. The last (first) layer of the encoder (decoder) network is a dense network with two (one) layers (see Fig.~\ref{fig:12c}). In the case of the encoder, a hierarchical (conditional) relationship among variables is implemented as described in Fig.~\ref{fig:7a}. The convolutional networks are implemented as a simple sequence of five gated convolutions, whose detailed implementation is given in Fig.~\ref{fig:12c}. Notice the use of batch normalization and dropout. The latter was only used in the decoder, to prevent over-fitting, with a drop-rate of $0.2$.

We trained our models using batches of size $100$, and the Adam optimizer with an initial learning rate of $3 e^{-3}$, exponentially decaying to a minimum learning rate of $1 e^{-4}$ after $1800$ epochs. The temperature parameter $\tau$ defined in Eq.~\ref{eq:smooth} for the Gumbel trick is typically annealed from large to small values. We however did not find a real advantage in doing so, and we fixed the parameter to the low value $\tau = 1/7$ throughout the training. To improve training and avoid collapse of the approximating posterior to trivial local minima, we have linearly annealed the KL term from zero to its full value within $200$ epochs.

In general we have trained our models using an importance-weighted estimate of the likelihood. As first described in Ref.~\cite{burda2015importance}, a $K$-sample weighting estimate of the log-likelihood can be written as:
\bea
\mathcal L_K = \E_{\bzeta_1, \dots,\z \sim q_{\bphi}(\z|\x)}\left[ \log \frac1K \sum_{k=1}^K\frac{p_{\btheta}(\z, \x)}{q_{\bphi}(\z|\x)}\right]\,,
\label{eq:k-log}
\eea
which is equivalent to the ELBO defined in Eq.~\ref{ELBOKL} for $K=1$ and converges to the exact log-likelihood for $K \rightarrow \infty$. We also found useful, to reduce the variance of the gradients of $\mathcal L_K$, to use a multi sample evaluation of the gradients per data point $\x$. In other words, we can use the following for training:
 \bea
\mathcal L_{K,D} = \E_{\z_{k,d} \sim q_{\bphi}(\z|\x)}\left[ \log \frac1K \sum_{k=1}^K\frac{p_{\z}(\z_{k,d}, \x)}{q_{\bphi}(\z_{k,d}|\x)}\right]\,,
\label{eq:k-d-log}
\eea
with $k=1,\dots,K$ and $d=1,\dots,D$. Notice that Eq.~\ref{eq:k-d-log} requires sampling  $K D$ latent configurations per data-point $\x$. This can be parallelized on GPU by effectively working, in our case, with batches of size $K D \times 100$. In our experiments we found it effective to have $K D = 8$ and to change the relative values of $K$ and $D$ while keeping their product constant. Every $200$ epochs we changed their value as follows: $(K, D) = (1, 8) \rightarrow (2, 4) \rightarrow (2, 4) \rightarrow (4, 2) \rightarrow (4, 2) \rightarrow (8, 1)$ and kept it constant afterwards. While a larger $K$ results in a tighter variational lower bound, it also makes harder training the approximating posterior, the reason being that in the limit of large $K$ the bound $\mathcal L_K$ does not depend on $q_{\bphi}(\bzeta|\x)$. We found this $K\leftrightarrow D$ anneal to be more efficient at both training the approximating posterior and training on a tighter bound to the log-likelihood. We used the same technique, with $K=1000, D=1$ as the estimate of the log-likelihood.

%%%%%%%%%%%%%%%%%%%%%%%%%%%%%%%%%%%%%
\section{Sample collection with D-Wave 2000Q}
\label{sec:exp}
%%%%%%%%%%%%%%%%%%%%%%%%%%%%%%%%%%%%%
To estimate the negative phase with D-Wave annealers, we used $1000$ samples obtained with independent annealing runs. For each gradient evaluation, we 
performed $5$ random spin-reversal transformations and collected $200$ samples each time. We used a forward annealing schedule with a $1\:\mu s$ forward anneal up to $s=0.5$, where we paused for $10\: \mu s$. After the pause we performed a $10\: ns$ quench to finish the anneal. After a bit of experimentation, we found this particular annealing schedule to slightly improve training, although a simple forward annealing without pause-and-quench also worked well. We did not perform any post-processing of the samples, which we used as-is to compute the negative phase.

An important question for future works is whether a more careful choice of the annealing schedule, possibly with longer pauses, can stabilize the effective temperature at which samples are drawn from the hardware. We discuss the importance of this aspect in the next section.

%%%%%%%%%%%%%%%%%%%%%%%%%%%%%%%%%%%%%
\section{Estimating $\beta^{*}_{eff}$ during training}
\label{sec:betaest}
%%%%%%%%%%%%%%%%%%%%%%%%%%%%%%%%%%%%%
\begin{figure}[h]
\begin{center}
\includegraphics[width=1\columnwidth]{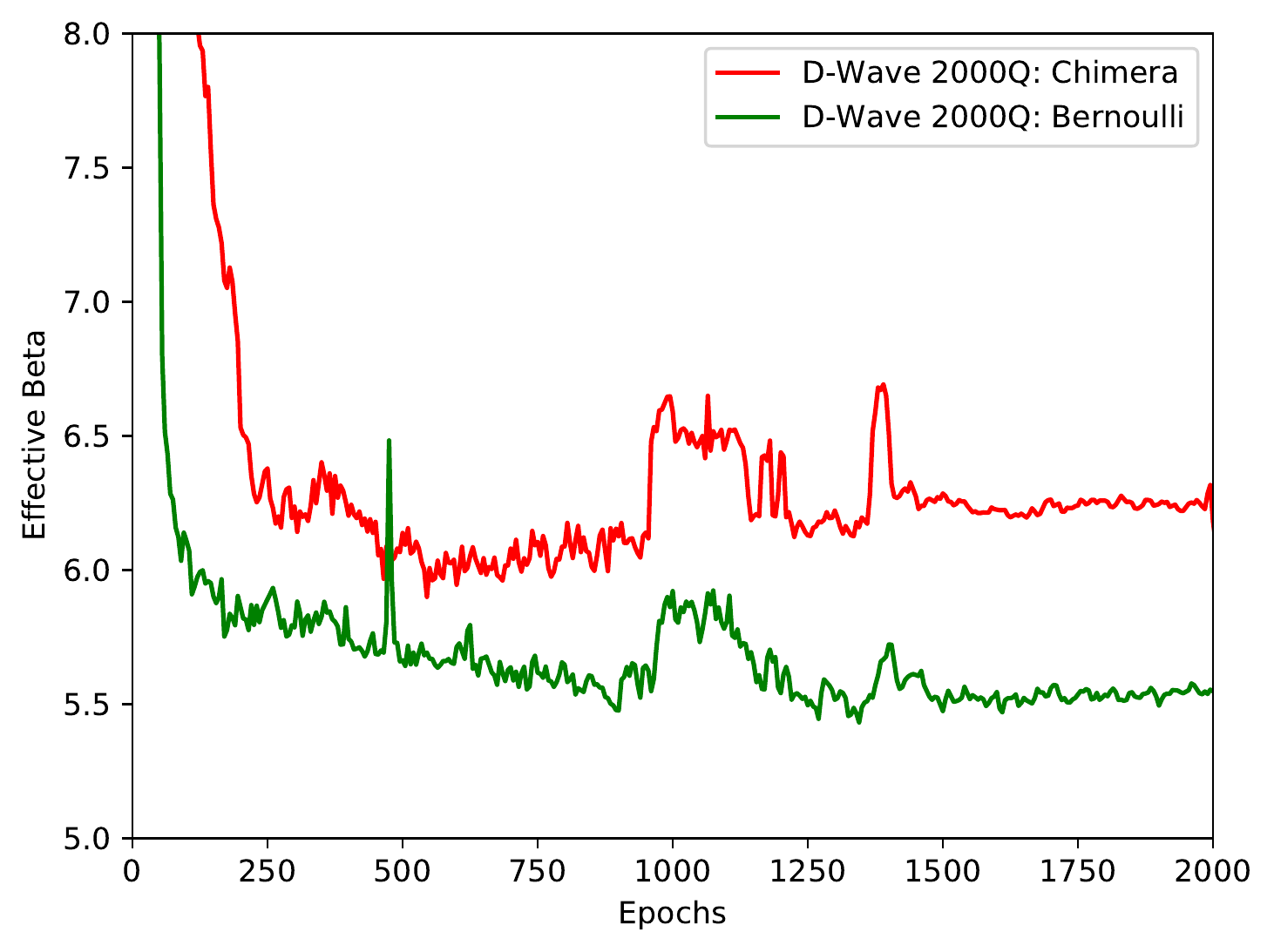}
\caption{$\beta^{*}_{eff}$ evaluated on two simultaneous runs on a D-Wave 2000Q (Chimera and Bernoulli priors). The fluctuations of its value on the two runs are correlated, indicating our evaluation of $\beta^{*}_{eff}$ is effectively probing fluctuations of the physical temperature of the device.}
\label{fig:13}
\end{center}
\end{figure}
As noticed in Ref.~\cite{benedetti2016estimation}, training a BM with a quantum annealer does not necessarily require the knowledge of the effective sampling temperature introduced in Eq.~\ref{eq:freez}. Indeed, $\beta^{*}_{eff}$ can be absorbed into the learning rate $\gamma$:
\bea
\pd \log \tilde p_{b, W}(\z) & = &  \gamma  \left( - \pd \Ham_{b, W}(\z)  + \E_{\bar \z \sim p_{b, W}}[\pd \Ham_{b, W}(\bar \z)]\right)= \nn 
& = &  \gamma' \left(- \pd \Ham_{h, J}(\z)  + \E_{\bar \z \sim p_{b, W}}[\pd\Ham_{h, J}(\bar \z)]\right) \nn
\gamma' &=& \gamma \beta^{*}_{eff}\,.
\label{eq:GT3}
\eea
While this is still true for the gradients of the parameters of the RBM placed in the latent space of a VAE, correctly evaluating the gradients of the inference parameters $\bphi$ requires knowledge of $\beta^{*}_{eff}$. To see this it suffices to note that the samples in the positive phase depends on the inference parameters through the reparameterization trick. During training: $\z \rightarrow \bzeta(\bphi, \brho)$. Tracking where these gradients come from, we have:
\bea
\gamma \pd_{\bphi} (ELBO) & = & - \gamma \pd_{\bphi} \log q_\bphi(\bzeta(\bphi, \brho)|\x) + \nn 
 &  & - \gamma \beta^{*}_{eff} \pd_{\bphi} \Ham_{h, J}(\bzeta(\bphi, \brho)\,,
\eea
so that the correct evaluation of the gradients with respect to the inference parameters requires the (approximate) knowledge of the effective temperature $\beta^{*}_{eff}$.

In our experiments, we have performed a real-time estimation of  $\beta^{*}_{eff}$, which we used as in the equation above to correctly estimate the gradients for the inference parameters. To do so, we employed an auxiliary BM that we trained in parallel with the VAE on the negative samples obtained by the quantum annealers. The parameters of the BM are shared according to Eq.~\ref{eq:freez}, with the only trainable parameter being $\beta^{*}_{eff}$. In other words, we update $\beta^{*}_{eff}$ as follows:
\bea
\beta^{*}_{eff} & \rightarrow&  \beta^{*}_{eff} + \nn
& + & \gamma \left( - \E_{\bar \bar\z \sim p^{HW}_{h, J}}[\Ham_{h, J}(\z)]  + \E_{\bar \z \sim p^{BM}_{b, W}}[\Ham_{h, J}(\bar \z)]\right)\,,\nn
\label{eq:temp_grad}
\eea
where the first expectation is evaluated with the hardware samples; the second, with thermal samples from the auxiliary BM (obtained with PA).

In Fig.~\ref{fig:13} we show the value of $\beta^{*}_{eff}$ estimated with the method above on two simultaneous runs on a D-Wave 2000Q. Its value typically drops while the KL term is annealed (200 epochs in our experiments), and subsequently stabilizes. Some fluctuations are correlated among independent runs, and are related to real fluctuations of the physical temperature of the device.

We have noticed that, due the use of the KL anneal and the presence of a non-negligible change in $\beta^{*}_{eff}$ during training, using a time-dependent evaluation of the effective temperature is important to stabilize training. While computing a single gradient as in Eq.~\ref{eq:temp_grad} is much more robust than training all the weights of a comparable BM, the method is not completely scalable and requires thermal sampling with classical algorithms. It will be critical, in future works, to implement training procedures with stable values of $\beta^{*}_{eff}$, which could be kept constant, using values predetermined by previous experiments or simply treated as a hyper parameter whose value must be appropriately fixed. Eventually, the use of more advanced annealing schedules, with longer pauses and more carefully chosen pause-points, should allow a direct connection between $\beta^{*}_{eff}$ and the physical temperature of the annealer, thus removing the necessity of learning $\beta^{*}_{eff}$ from experiments. 

%%%%%%%%%%%%%%%%%%%%%%%%%%%%%%%%%%%%%
\section{Chimera and Pegasus connectivities}
\label{sec:CHI-PEG}
%%%%%%%%%%%%%%%%%%%%%%%%%%%%%%%%%%%%%

\begin{figure*}[h]
\begin{center}
\subfigure[\, Chimera architecture with 288 units.]{\includegraphics[width=0.9\columnwidth]{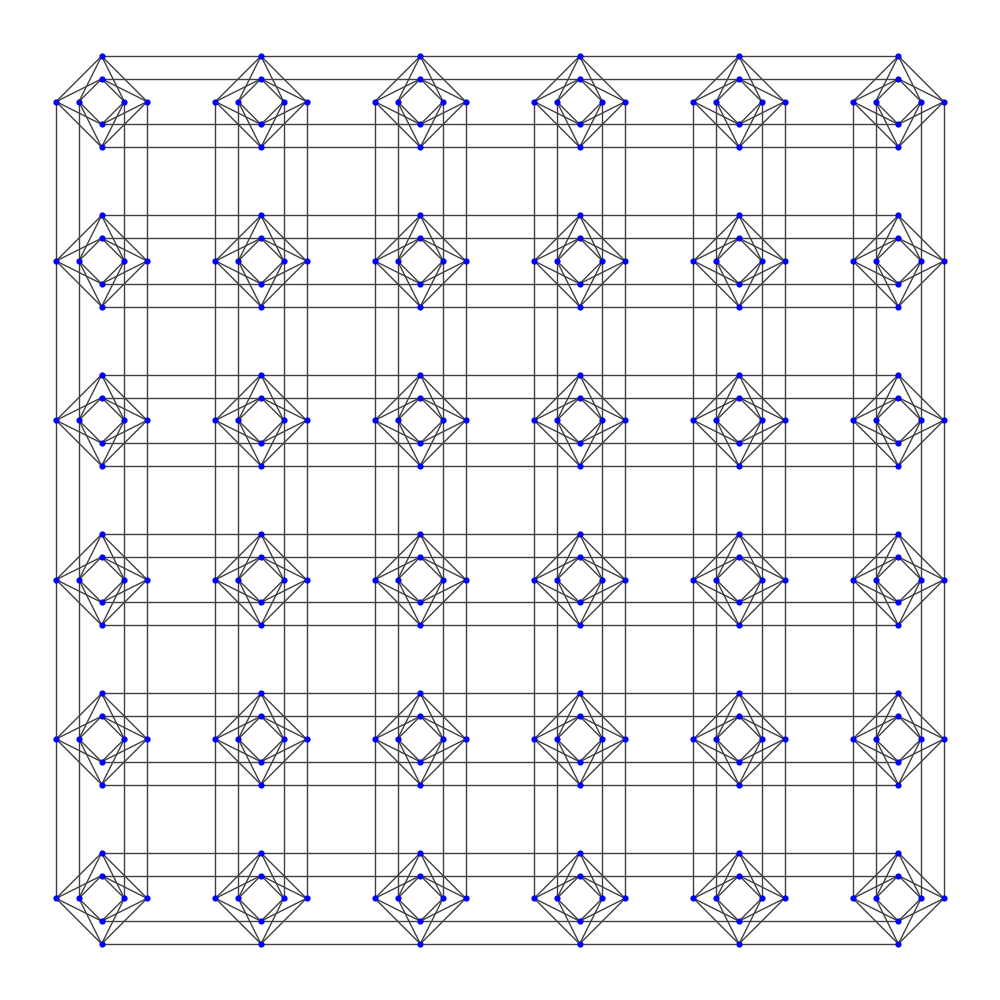}\label{fig:14a}}
\subfigure[\, Pegasus architecture with 288 units.]{\includegraphics[width=0.9\columnwidth]{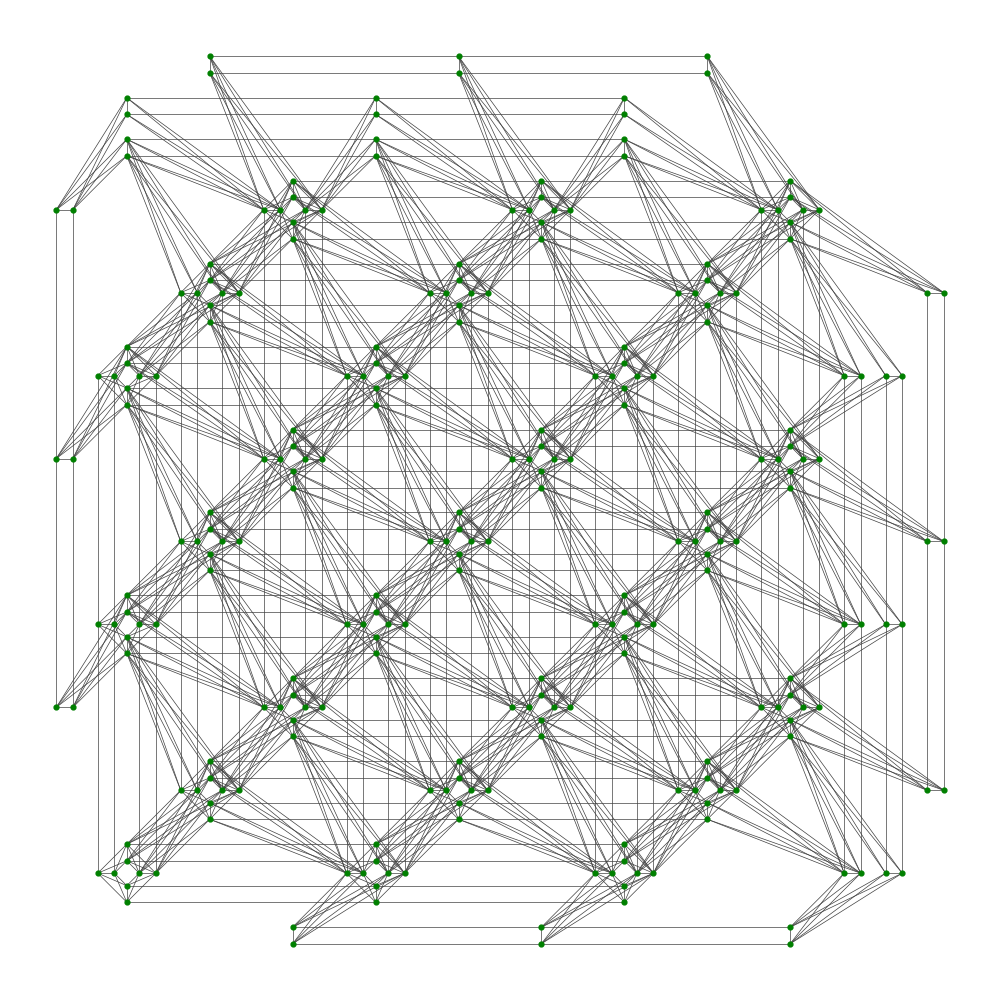}\label{fig:14b}}
\subfigure[\, Chimera cells.  ]{\includegraphics[width=0.8\columnwidth]{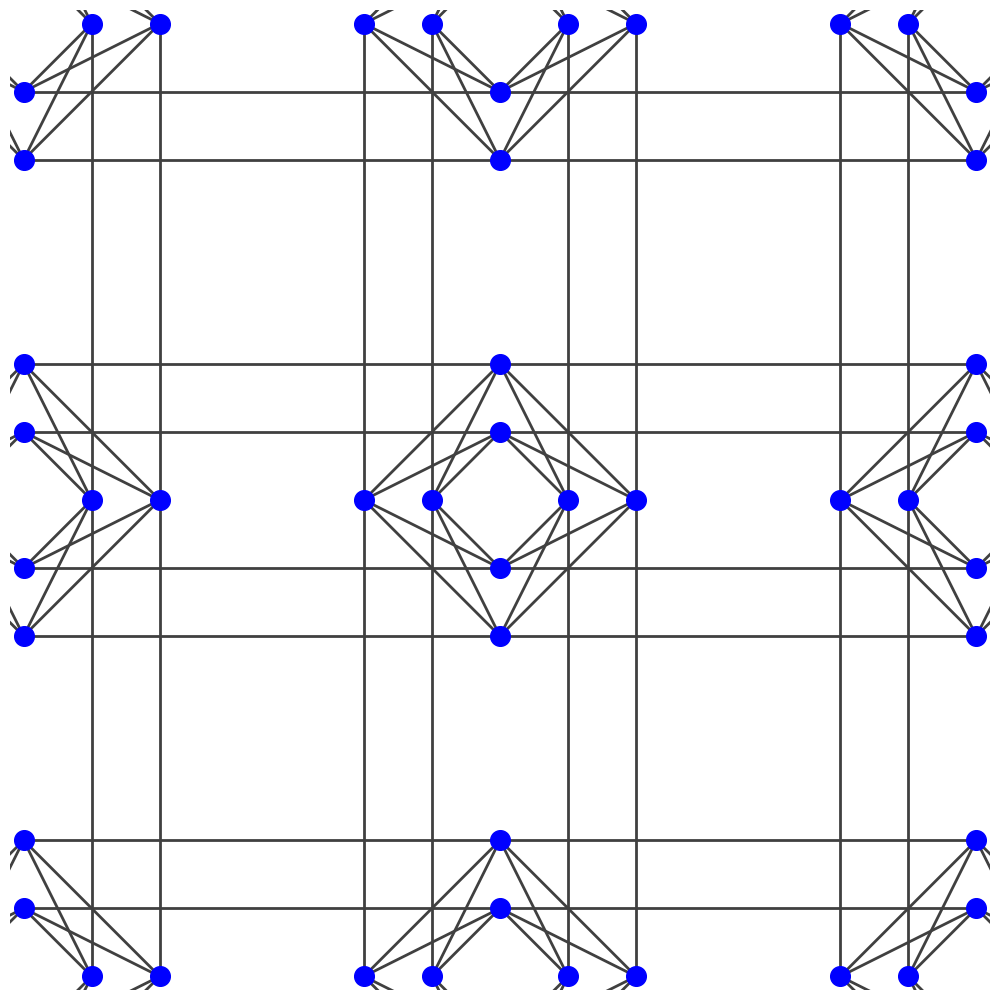}\label{fig:14c}} \quad \quad
\subfigure[\, Pegasus cells.]{\includegraphics[width=0.8\columnwidth]{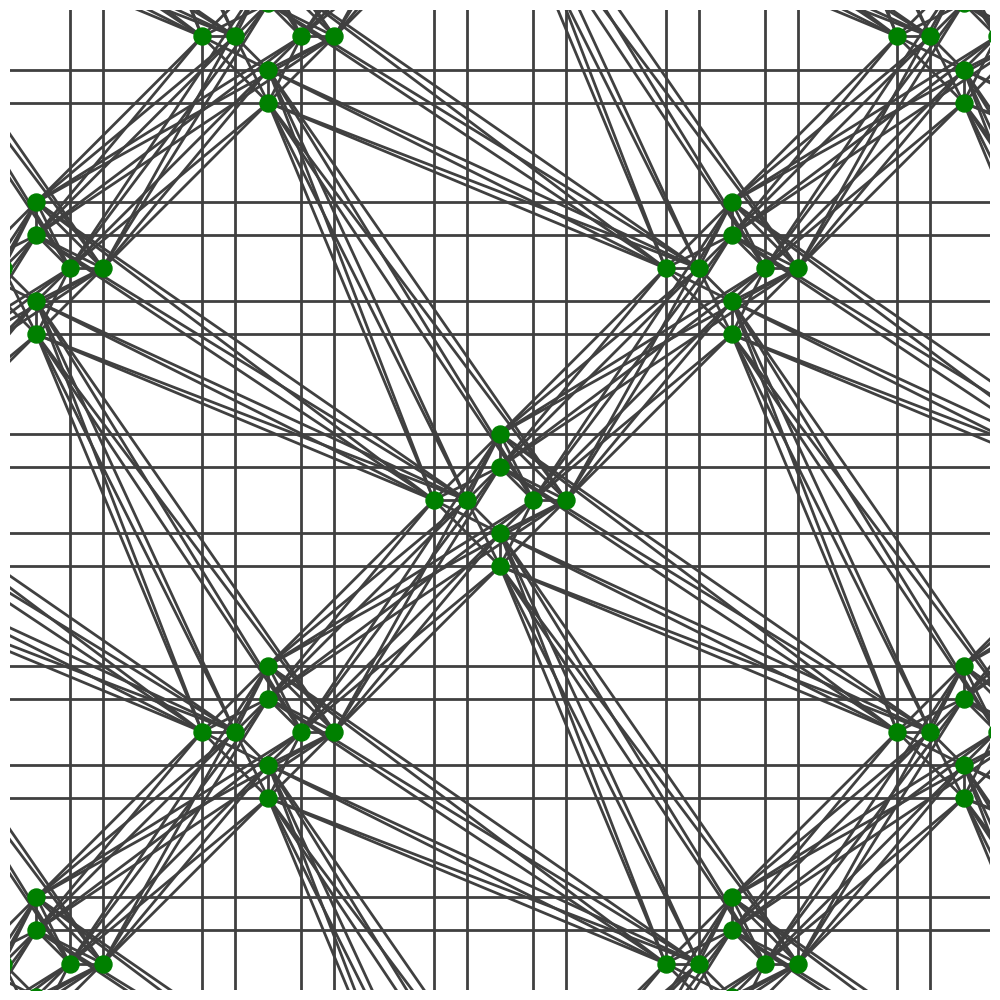}\label{fig:14d}}

\caption{Physical connectivities used in this work.}
\label{fig:14}
\end{center}
\end{figure*}

In Fig.~\ref{fig:14} we show the Chimera and Pegasus connectivities on 288 qubits used in all the experiments performed in this work. The Chimera graph (Fig.~\ref{fig:14a}) is a bipartite, two-dimensional tiling of a unit cell (Fig.~\ref{fig:14c}) with 8 qubits. The Pegasus graph (Fig.~\ref{fig:14b}) is a quadri-partite, two-dimensional tiling of a unit cell (Fig.~\ref{fig:14d}) with 8 qubits~\cite{boothby2019next}.

\end{document}